\documentclass[superscriptaddress,orcidlink,
 reprint,
 nofootinbib,
 amsmath,amssymb,
 aps,
]{revtex4-1}
\usepackage{orcidlink}
\usepackage{graphicx}% Include figure files
\usepackage{dcolumn}% Align table columns on decimal point
\usepackage{bm}% bold math
\usepackage{amsmath}
\usepackage{enumitem}
\usepackage{subcaption}
\usepackage{float}
\usepackage{ulem}
\usepackage{verbatim}
\begin{document}

\title{Testing the consistency of early and late cosmological parameters with BAO and CMB data}

\author{Guanlin Liu \orcidlink{0009-0000-9225-7169}}
\email{irithyll0110@mail.ustc.edu.cn}
\affiliation{CAS Key Laboratory for Researches in Galaxies and Cosmology, Department of Astronomy, University of Science and Technology of China, Chinese Academy of Sciences, Hefei, Anhui 230026, China}
\affiliation{School of Astronomy and Space Sciences, University of Science and Technology of China, Hefei 230026, China}

\author{Yu Wang}
\affiliation{CAS Key Laboratory for Researches in Galaxies and Cosmology, Department of Astronomy, University of Science and Technology of China, Chinese Academy of Sciences, Hefei, Anhui 230026, China}
\affiliation{School of Astronomy and Space Sciences, University of Science and Technology of China, Hefei 230026, China}

\author{Wen Zhao \orcidlink{0000-0002-1330-2329}}
\email{wzhao7@ustc.edu.cn}
\affiliation{CAS Key Laboratory for Researches in Galaxies and Cosmology, Department of Astronomy, University of Science and Technology of China, Chinese Academy of Sciences, Hefei, Anhui 230026, China}
\affiliation{School of Astronomy and Space Sciences, University of Science and Technology of China, Hefei 230026, China}

%\title{Constraining the early and late cosmological parameters at the same time to examine $\Lambda$CDM and determine the Hubble tension}

%[0009-0000-9225-7169]
%[0000-0002-1330-2329]

\begin{abstract}
The recent local measurements of the Hubble constant $H_0$, indicate a significant discrepancy of over 5$\sigma$ compared to the value inferred from \textit{Planck} observations of the cosmic microwave background (CMB). In this paper, we try to test the standard cosmological model $\Lambda$CDM by testing the consistency of early and late cosmological parameters in the same observed data. In practice, we simultaneously derive the early and late parameters using baryon acoustic oscillation (BAO) measurements, which provide both low and high-redshift information. CMB data are also included in the analysis as a "distance prior", which tracks the same BAO feature and resolve parameter degeneracy. By using the parameter $\omega_m = \Omega_mh^2$, we introduce ${\rm ratio}(\omega_m)$, defined as the ratio of $\omega_m$ which are constrained from high and low-redshift measurements respectively, to quantify the consistency between early and late parameters. We obtained a value of ${\rm ratio}(\omega_m) = 1.0069\pm0.0070$, indicating there is no tension between early parameters and late parameters in the framework of $\Lambda$CDM model. As a result, our test does not expose the defects of the $\Lambda$CDM model. In addition, we forecast the future BAO measurements of ${\rm ratio}(\omega_m)$, using several galaxy redshift surveys and 21 cm intensity mapping surveys, and find that these measurements can significantly improve the precision of cosmological parameters.
\end{abstract}

\maketitle

\section{Introduction} \label{sec1} % background: Hubble tension + modification
The standard cosmological model, $\Lambda$CDM, has withstood the rigorous scrutiny due to significant advances in observations of the cosmic microwave background (CMB) \cite{2020planck,balkenholspt,aiola2020act}, Type Ia supernovae (SNIa) \cite{scolnic2018pantheon,brout2022pantheon+} and large-scale structure \cite{alam2017sdss3,alam2021completed,abbott2018desy1} in recent years. However, tensions have emerged for several parameters when they are inferred from distinct observations within the $\Lambda$CDM. The most significant tension is related to determinations of the Hubble constant $H_0$. The representative difference is between the SH0ES \footnote{Supernovae and $H_0$ for the Equation of State.} \cite{SH0ES2022} determination of $H_0$ and the value inferred from \textit{Planck} \cite{2020planck} for the $\Lambda$CDM cosmology, which has increased to $5\sigma$ level \cite{verde2019tensionof5sig,abdalla2022cosmologyreview}. Actually, the tension exists in not only these two values, but also in two sets of measurements \cite{abdalla2022cosmologyreview}. 

One set contains indirect estimates at early times which are cosmological model dependent, such as CMB and Baryon Acoustic Oscillation (BAO) \cite{cuceu2019bao+bbn1,schoneberg2022bao+bbn2}. As mentioned above, the \textit{Planck} 2018 release \cite{2020planck} predicts $ H_0 = 67.36 \pm 0.54 \ \rm km \ s^{-1} \ Mpc^{-1} $ in combination with the CMB lensing. Moreover, the final combination of the BAO measurements from galaxies, quasars, and Lyman-$\alpha$ forest (Ly-$\alpha$) from Baryon Oscillation Spectroscopic Survey (BOSS) prefers $ H_0 = 67.64^{+0.94}_{-1.03} \ \rm km \ s^{-1} \ Mpc^{-1} $ \cite{schoneberg2022bao+bbn2} involving a prior on the physical density of the baryons $\Omega_bh^2$ derived from the latest Big Bang Nucleosynthesis (BBN) measurements of the primordial deuterium \cite{cooke2018bbn1,mossa2020bbn2}. In contrast, another set includes direct late time cosmological model-independent measurements, such as distance ladders \cite{SH0ES2022,jang2017trgb1,freedman2020trgb2,huang2020miras,kourkchi2020tully-fisher,2017PhRvD..96j3529S} and strong lensing \cite{wong2020h0licow}. The determination of $H_0$ is based on calibration of the SNIa using the Cepheids, Miras and Tip of the Red Giant Branch (TRGB), which prefers a higher value. The latest SH0ES measurement gives $H_0 = 73.04 \pm 1.04 \ \rm km \ s^{-1} \ Mpc^{-1}$ using Cepheid period–luminosity relations to calibrate distances to Type Ia SNe host galaxies \cite{SH0ES2022}.

There are various potential solutions proposed for the Hubble tension. One kind of viewpoint thinks that some unknown systematic errors affect either early-universe or late-universe observation \cite{di2021systemerror}. %The Hubble tension can also be resolved by abandoning the cosmological principle \cite{abdalla2022cosmologyreview}, namely the assumption of a homogeneous and isotropic universe. %
However, the other viewpoint is related to the new physics beyond the standard cosmological model. The late time modifications of the $\Lambda$CDM expansion history have been showed that they cannot resolve the Hubble tension through a model independent method in \cite{lemos2019inverseladder,efstathiou2021inverseladder2}. Therefore, many solutions to the Hubble tension is to introduce new physics about early universe that increases the value of the Hubble constant inferred from the CMB.  An popular early-time resolution \cite{karwal2016ede1,mortsell2018ede2,poulin2019ede3} is an exotic early dark energy (EDE) that behaves like a cosmological constant at early time and dilutes faster than radiation at later time. It addresses the Hubble tension by decreasing the sound horizon while leaving the later evolution of the Universe unchanged.

% introduce our work and compare with other works

In order to understand the Hubble tension, it is important to ask the question: Whether or not the $\Lambda$CDM cosmology is free of defects? On the one hand, the authors of Ref.\cite{krishnan2021running,colgain2022revealing,malekjani2023negative} highlight that $H_0$ should pick up redshift if the $\Lambda$CDM model is breaking down. On the other hand, our goal is testing whether the constant parameter $\omega_m = \Omega_mh^2$ of $\Lambda$CDM changes with redshift. If we find $\omega_m$ derived from different redshift measurements are not consistent, then we can conclude the $\Lambda$CDM model fails the test and is necessary to be modified. For instance, the pressureless matter assumption may not be supported. In this paper, we achieve the simultaneous constraints on the early and late cosmological parameters $\omega_m$ with the help of BAO measurements that contain both low and high-redshift universe information. In practice, two distinct $\Lambda$CDM cosmological parameter sets are employed to construct the expansion history of the early and late universe separately. However, due to the degeneracy, CMB data is combined with BAO data for breaking it and $H_0$ is replaced by a new parameter ${\rm ratio}(\omega_m)$, which will be introduced detail in Section \ref{sec2}. As a result, we find the consistency of early and late cosmological parameter protects the standard cosmological model. In other words,there is no evidence of new physics beyond $\Lambda$CDM cosmology.

%The analysis presented here resembles studies that give constraints on early or late parameters separately. The authors of \cite{lemos2023cmbearly,vonlanthen2010cmbearly2,audren2012cmbearly3} analyze CMB data in a manner which is as independent as possible of the model of late-time cosmology and obtain robust constraints on early cosmology parameters. The late cosmological parameters are regarded as cosmological parameters constrained by low redshift measurements. For example, Gaussian Process (GP) regression \cite{holsclaw2011gpr1,shafieloo2012gpr2,seikel2012gpr3} is the most commonly used technique to reconstruct the Hubble parameter $H(z)$ directly from data without assuming a cosmological model. It means $H_0$ can be determined by simply extrapolating $H(z)$ to $z = 0$ \cite{verde2017late1,gomez2018late2,moresco2022late3}. Our results will be compared to those mentioned above. We also follow methods developed by \cite{philcox2022rdagnostic} to test EDE in a different way.

This article is organized as follows. In Section \ref{sec2}, we introduce the methods and datasets used in this analysis. The results and related discussions are presented in Section \ref{sec3}. We simulate future BAO experiments and analyze the results in Section \ref{sec4}. Finally, Section \ref{sec5} presents our conclusions in this article.

\section{Methodology} \label{sec2}
In the standard Friedmann-Lemaitre-Robertson-Walker (FLRW) universe, the Hubble parameter $H(z)$ on the basis of flat $\Lambda$CDM is given by:
\begin{equation}\label{eq-1}
    H(z)=H_0\sqrt{{\Omega}_r(1+z)^4+{\Omega}_m(1+z)^3+{\Omega}_{\Lambda}},
\end{equation}
where ${\Omega}_r$, ${\Omega}_m$ and ${\Omega}_{\Lambda}$ are the fractional densities of radiation, matter and dark energy at redshift $ z = 0 $ and constrained by $ {\Omega}_r+{\Omega}_m+{\Omega}_{\Lambda}=1 $. %$E(z)$ is defined to represent components of the Universe: $E(z) \equiv H(z)/H_0$.
% 3 portions: revise model motivation; introduce model; use 
% motivation: tension prompt "E-L LCDM"; analyze model-based:CMB+BAO; test of LCDM
% category to illustrate clearly (3 parts)

As previously indicated regarding the Hubble tension, the 5$\sigma$ difference between early time and late time inspires us to investigate whether its reason isfrom $\Lambda$CDM model. It will be completed through a test of the cosmological parameter consistency. In practice, the BAO data which contains both low and high-redshift universe information will be analyzed in the framework of the flat $\Lambda$CDM cosmology. We employ two distinct independent cosmological parameter sets to describe the expansion history of the early and late universe separately. Therefore, the consistency of these two cosmological parameter sets can help us distinguish the potential time-dependence of cosmological parameter. In order to realize it, the parameters in the standard cosmological model can be generalized as follows:
\begin{enumerate}[label=(\roman*),align=left]
    \item Parameter set undergoes a transition from \\ $\lbrace H_0, \Omega_m \rbrace$ to $\lbrace \tilde{H}_0, \tilde{\Omega}_m, H_0, \Omega_m \rbrace$. Note that, here and afterwards, the parameters with $\emph{tilde}$ label that their values are derived from early cosmological information and we call them as ``early cosmological parameters” in this article, while the parameter without $\emph{tilde}$ label that their values are derived from late cosmological information and we call them as ``late cosmological parameters”.

    \item Recombination redshift $z_*$ serves as the dividing line between the early and late universe. Namely the early parameters $\left( \tilde{H}_0, \tilde{\Omega}_m \right) $ are determined from the measurements whose redshift is larger than the recombination redshift $z_*$, while the late parameters remain similar. Therefore we realize constructing the early and late expanding history of the Universe individually.
    
    \item In order to test $\Lambda$CDM model through the constraint of both early and late parameters simultaneously, we can compare whether early parameters $\left( \tilde{H}_0,\tilde{\Omega}_m\right)$ and $\left( H_0,\Omega_m\right)$ are consistent. We are able to determine whether the tension between early and late parameters of $\Lambda$CDM exists by the combination of BAO and CMB data. 
\end{enumerate}

More details will be discussed in the data analysis part. 
%\begin{equation}\label{eq-3}
    %\Gamma(z)=\int_{0}^{z}\frac{\mathrm{d}z'}{E(z')},
%\end{equation}
%and 
%\begin{align}\label{eq-4}
%S_k(x) =
%\begin{cases}
%  \sin{\sqrt{-\Omega_k}x}/\sqrt{-\Omega_k} & \Omega_k < 0, \\
%  x & \Omega_k = 0, \\
%  \sinh{\sqrt{\Omega_k}x}/\sqrt{\Omega_k} & \Omega_k > 0.
%\end{cases}
%\end{align}

\subsection{BAO data} % BAO introduction; measurement ; data origin ; likelihood
In the primordial plasma of the early universe, photon-matter interactions give rise to substantial heat and outward pressure. This pressure drives baryons and photons out of the overdensity area at a speed of sound $c_s$. As a result, the spherical sound waves propagate baryon density fluctuations in the Universe until the recombination epoch. The distance traveled by sound waves between the end of inflation and the decoupling of baryons from photons after recombination is referred to as the sound horizon $r_s$. Dark matter remains at the center of acoustic wave since it is influenced solely by gravity. Finally, the baryons and dark matter resulted in a structure that included matter overdensities both at the original site and at the sound horizon scale. Thus, we can anticipate observing a more significant number of galaxy pairs separated by the sound horizon distance than other length scales. In other words, the BAO method relies on the imprint left by propagation of acoustic density waves to provide a cosmic standard ruler which has a comoving scale $r_s$.
%It represents the distance traveled by sound waves between the end of inflation and the decoupling of baryons from photons after recombination. 

The comoving sound horizon $r_s(z)$ is given by \cite{eisenstein1998baorigin}:
\begin{eqnarray}\label{eq-5}
    r_s(z)=\frac{c}{H_0}\int_{z}^{\infty}\frac{c_s}{E(z')}\mathrm{d}z' ,
\end{eqnarray}
where $c$ is the speed of light, the sound speed is $c_s=1/\sqrt{3(1+R_ba)}$, with $R_ba=3\rho_b/(4\rho_r)$, and $R_b=31500\omega_b(T_{CMB}/2.7K)^{-4}$ with $\omega_b=\Omega_bh^2$ and the CMB temperature $T_{CMB} = 2.7255K$. It is necessary to emphasize that the radiation component cannot be ignored when calculating the comoving sound horizon $r_s$. It can be described by the matter fraction $\omega_m$ applying matter-radiation equality relation $\Omega_r = \Omega_m/(1+z_{eq})$, and $z_{eq} = 2.5\times 10^4 \omega_m (T_{CMB}/2.7K)^{-4}$. %We assume the CMB temperature $T_{CMB} = 2.7255K$.

The BAO feature appears in both the line-of-sight direction and the transverse direction and provides a measurement of $D_H(z)/r_d$ and $D_M(z)/r_d$ individually:
\begin{eqnarray}
    \frac{D_H(z)}{r_d} &=& \frac{c}{H(z)r_d},\label{eq-6} \\
    \frac{D_M(z)}{r_d} &=& \frac{c}{H_0r_d}\Gamma(z), \label{eq-7}
\end{eqnarray}
where $r_d$ is the sound horizon at the drag epoch $r_s(z_d)$ and $D_M$ is the same as transverse comoving distance \cite{hogg1999distance}. The transverse comoving distance, which depends on FLRW metric to an object at redshift $z$, is given by 
\begin{equation}\label{eq-2}
    D_M(z)=r(z)=\frac{c}{H_0}S_k(\Gamma(z)),
\end{equation}
where $\Gamma(z)=\int_{0}^{z}{\mathrm{d}z'}/{E(z')}$, $E(z)=H(z)/H_0$. Note that the angular diameter distance $D_A(z)$ has relation with the comoving angular diameter distance $r(z)$: $D_A(z)=r(z)/(1+z)$. Because $S_k(x)=x$ for flat $\Lambda$CDM model, we have $r(z) = c\Gamma(z)/H_0$ in the following analysis.

The BAO measurements were also historically summarized by a single quantity representing the spherically averaged distance
\begin{equation} \label{eq-8}
    D_V(z) \equiv [zD_M^2(z)D_H(z)]^{1/3}.
\end{equation}

In the Appendix E of Ref.\cite{hu1996analyticalformula}, the authors develop fitting functions for convenience which are valid at the percent level for an extended range of parameter space, $0.0025\le \omega_b \le0.25$ and $0.025\le \omega_m \le0.64$. Therefore, the fitting functions of drag epoch redshift $z_d$ and recombination redshift $z_*$ are widely used in relevant works \cite{komatsu2009five,wang2007distanceprior-1,wang2013distanceprior-2,wang2019distanceprior-3}. We also introduce these fitting functions to calculate the sound horizon. The redshift of the drag epoch can be approximated as \cite{hu1996analyticalformula}
\begin{equation}\label{eq-9}
    z_d=\frac{1345\omega_m^{0.251}}{1+0.659\omega_m^{0.828}}[1+b_1\omega_b^{b_2}],
\end{equation}
where
\begin{equation}\label{eq-10}
    b_1=0.313\omega_m^{-0.419}[1+0.607\omega_m^{0.674}], b_2=0.238\omega_m^{0.223}.
\end{equation}

It is obvious that BAO measurements provide simultaneous information about the early and late universe through the comoving sound horizon $r_d$ and Hubble parameter $H(z)$, which enables us to determine the physics in the early and late universe simultaneously.

To calculate the sound horizon $r_d$ and complete the set of cosmological parameters, we must also include information on baryon density $\Omega_b$ as an early parameter not only because it derives from CMB observation, but also it is just entailed to compute sound horizon $r_d$. Carrying over the remaining parameters introduced above, the initial full parameter set is now: 
\begin{align*}
\centering
\lbrace \tilde{\Omega}_b,\tilde{\Omega}_m,\tilde{H}_0,\Omega_m,H_0 \rbrace .
\end{align*}

Then, the BAO measurements can be expressed in this parameter set:
\begin{equation}
\frac{D_H(z)}{r_d} = \frac{\left[(1+z)^3+\Omega_l-1 \right]^{-\frac{1}{2}}}{\int_{z_d}^{\infty}\left[3\left(1+\frac{R_b}{1+z'}\right)(1+z')^3 f(\tilde{\omega}_m,\omega_m)\right]^{-\frac{1}{2}} \mathrm{d}z'}, \label{eq-22}
\end{equation}
where $\Omega_l \equiv 1/\Omega_m$, and
\begin{equation}
    f(\tilde{\omega}_m,\omega_m) \equiv \frac{\tilde{\omega}_m}{\omega_m} \left(1+\frac{1+z}{1+z_{eq}} \right). \label{eq-add1}
\end{equation}

The Hubble constant $H_0$ and matter density fraction $\Omega_m$ are fully degenerate with the form $\omega_m$. We can also find in Eq.\eqref{eq-add1} the ratio of $\omega_m$ is degenerate with the late cosmological parameter $\omega_m$ in the form of a sum due to $z_{eq}\propto \tilde{\omega}_m$. We will see the CMB data can break this degenerate relation in the next section. In addition, constraining this ratio help us compare the early and late cosmological parameters directly. Therefore, we introduce ${\rm ratio}(\omega_m) \equiv \tilde{\omega}_m/\omega_m$ as a new parameter. The analysis of transverse measurement $D_M(z)/r_d$ is similar to $D_H(z)/r_d$.

The BAO measurements used in this work are taken from the 6dF Galaxy Survey (6dFGS,\cite{beutler20116df}), the complete Sloan Digital Sky Survey (SDSS) data release, including SDSS-Main Galaxy Sample (MGS,\cite{ross2015mgs}), the BOSS and the extended Baryon Oscillation Spectroscopic Survey (eBOSS) DR16 data \cite{alam2021completed,bouroux2020lya}. A specific description of BAO data is shown in TABLE \ref{tab:table1}.

The likelihood of BAO $\mathcal{L}_{BAO}$ have different forms for different datasets. 
The covariance matrix of SDSS DR12 and eBOSS Quasar data, and the full (non-Gaussian) likelihoods for the Ly$\alpha$ forest data \cite{bouroux2020lya} and emission line galaxies (ELG) analysis \cite{de2021elg}, are from the public SDSS repository\footnote{https://svn.sdss.org/public/data/eboss/DR16cosmo/tags\\/v1\_0\_0/likelihoods/.}.

\begin{table*}[htb]
\captionsetup{justification=raggedright}
\begin{ruledtabular}
\begin{tabular}{ccccc}
 $z_{eff}$ & Measured Quantity & Value & Dataset & Reference \\ \hline
    0.106 & $r_d/D_V$ & $0.336\pm0.015$ & 6dFGS & \cite{beutler20116df} \\
    0.15 & $D_V/r_d$ & $4.47\pm0.17$ & SDSS DR7 & \cite{ross2015mgs} \\
    0.38 & $D_H/r_d$ & $25.00\pm0.76$ & SDSS DR12 & \cite{alam2021completed} \\
    0.38 & $D_M/r_d$ & $10.23\pm0.17$ & SDSS DR12 & \cite{alam2021completed} \\
    0.51 & $D_H/r_d$ & $22.33\pm0.58$ & SDSS DR12 & \cite{alam2021completed} \\
    0.51 & $D_M/r_d$ & $13.36\pm0.21$ & SDSS DR12 & \cite{alam2021completed} \\
    0.70 & $D_H/r_d$ & $19.33\pm0.53$ & eBOSS LRG & \cite{alam2021completed} \\
    0.70 & $D_M/r_d$ & $17.86\pm0.33$ & eBOSS LRG & \cite{alam2021completed} \\
    0.85 & $D_V/r_d$ & $18.33^{+0.57}_{-0.62}$ & eBOSS ELG & \cite{alam2021completed} \\
    1.48 & $D_H/r_d$ & $13.26\pm0.55$ & eBOSS Quasar & \cite{alam2021completed} \\
    1.48 & $D_M/r_d$ & $30.69\pm0.80$ & eBOSS Quasar & \cite{alam2021completed} \\
    2.33 & $D_H/r_d$ & $8.93\pm0.28$ & eBOSS Ly$\alpha \times$ Ly$\alpha$ & \cite{bouroux2020lya} \\
    2.33 & $D_M/r_d$ & $37.6\pm1.9$ & eBOSS Ly$\alpha \times$ Ly$\alpha$ & \cite{bouroux2020lya} \\
    2.33 & $D_H/r_d$ & $9.08\pm0.34$ & eBOSS Ly$\alpha \times$ Quasar & \cite{bouroux2020lya} \\
    2.33 & $D_M/r_d$ & $37.3\pm1.7$ & eBOSS Ly$\alpha \times$ Quasar & \cite{bouroux2020lya} \\
\end{tabular}
\caption{\label{tab:table1}Datasets measuring the BAO peak that are used in our cosmological analysis. }
\end{ruledtabular}
\end{table*}

\subsection{CMB data} % simple CMB introduction ? likelihood introduction ?
CMB reflect the distance to the surface of last scattering $r(z_*)$ by the precisely measured locations and amplitudes of peaks of the acoustic oscillations. In practice, CMB measures two distance ratios: (i) the comoving distance to the surface of last scattering $r(z_*)$ divided by the sound horizon size at the recombination epoch, $r(z_*)/r_s(z_*)$, and (ii) the comoving distance to the surface of last scattering $r(z_*)$ divided by the comoving Hubble horizon size at the recombination epoch, $r(z_*)H(z_*)/c(1+z_*)$. The two ratios can be quantified and defined as the CMB distance prior $l_a$ and $R$ separately. The CMB temperature power spectrum in the transverse direction is characterized by the acoustic scale $l_a$, which determines the peaks spacing. Additionally, the shift parameter $R$ affects the CMB temperature spectrum along the line-of-sight direction, leading to the heights of the peaks. Therefore, CMB observation contribution in the constraint of cosmological parameter can be converted to CMB shift parameters for breaking degeneracy of parameters, instead of using the full CMB power spectrum\cite{wang2007distanceprior-1,wang2013distanceprior-2}:
\begin{eqnarray}
    R&=&\sqrt{\Omega_mH^2_0}r(z_*)/c, \label{eq-11} \\
    l_a&=&\pi r(z_*)/r_s(z_*). \label{eq-12}
\end{eqnarray}
These parameters can be considered as the CMB measurements which are extracted from CMB likelihood, where $z_*$ is the redshift of recombination. The sufficient information of \textit{Planck} data release \cite{2020planck} for constraint is given by CMB shift parameters with baryon density $\omega_b = \Omega_bh^2$.

The redshift $z_*$ can be calculated by the fitting formula \cite{hu1996analyticalformula}:
\begin{equation}\label{eq-13}
    z_*=1048[1+0.00124\omega^{-0.738}_b][1+g_1\omega^{g_2}_m],
\end{equation}
where
\begin{equation}\label{eq-9}
    g_1=\frac{0.0783\omega^{-0.238}_b}{1+39.5\omega^{0.763}_b}, g_2=\frac{0.560}{1+21.1\omega^{1.81}_b}.
\end{equation}
The final result, which is extracted from \textit{Planck} 2018 TT,TE,EE+lowE chains, can be transformed in terms of a data vector $\bm{v}=(R,l_a,\omega_b)^T$ \footnote{Actually, spectral index of the primordial power spectrum $n_s$ is entailed for completeness, however we neglect $n_s$ because it doesn't relate to our constraint.} and their covariance matrix. For a flat universe \cite{wang2019distanceprior-3}, this data vector is
\begin{equation}\label{eq-14}
\bm{v}=\begin{pmatrix}
    1.74963 \\
    301.80845 \\
    0.02237 \\
    \end{pmatrix}
\end{equation}
and the covariance matrix is
\begin{equation}\label{eq-15}
    \bm{C_v}=10^{-8}\times 
    \begin{pmatrix}    
        1598.9554 & 17112.007 & -36.311179  \\
        17112.007 & 811208.45 & -494.79813  \\
       -36.311179 & -494.79813& 2.1242182 
    \end{pmatrix} \nonumber
\end{equation}

The compression of CMB data forms the so-called distance priors \cite{wang2007distanceprior-1}, which can help us break the degeneracy of the BAO data:
\begin{equation}
    l_a = \frac{\pi R}{\int_{z_*}^{\infty}\left[3\left(1+\frac{R_b}{1+z'}\right)(1+z')^3 \left(1+\frac{1+z'}{1+z_{eq}} \right) \right]^{-\frac{1}{2}} \mathrm{d}z'}. \label{eq-20}
\end{equation}

Apparently, $\tilde{\omega}_m$ is the only cosmological parameter to be determined in the CMB shift parameter $l_a$. Namely, combining the CMB data can give a constraint on $\tilde{\omega}_m$ and break the degeneracy to release ${\rm ratio}(\omega_m)$. Therefore the final parameter set is now
\begin{align*}
\centering
\lbrace \tilde{\omega}_b,\tilde{\omega}_m,\Omega_m,{\rm ratio}(\omega_m) \rbrace .
\end{align*}

In summary, the baryon density $\tilde{\omega}_b$ and the matter density $\tilde{\omega}_m$ serve as early parameters to compute sound horizon $r_s$. $\Omega_m$ describes the evolution of late universe and ${\rm ratio}(\omega_m)$ is a bridge between them.

The loglikelihood function for CMB can be written in the form:
\begin{equation}\label{eq-21}
\mathcal{L}_{CMB} = -\frac{1}{2}\chi^2_{CMB} = -\frac{1}{2}(\bm{v}-\bm{v}_{th})^T \bm{C_v}^{-1} (\bm{v}-\bm{v}_{th}),
\end{equation}
and the complete likelihood for our cosmological analysis is the following $\mathcal{L}=\mathcal{L}_{\rm CMB}+\mathcal{L}_{\rm BAO}$. We run a Markov Chain Monte Carlo (MCMC) with the sampler \texttt{emcee} \cite{emcee} to constrain parameters, where the priors chosen for our Bayesian calculations regarding the cosmological parameters are uniform distributions: $\tilde{\omega}_b \in (0.001,0.3)$, $\tilde{\omega}_m \in (0.05,0.5)$, $\Omega_m \in (0.1,0.9)$, ${\rm ratio}(\omega_m) \in (0.1,10)$.

\section{Results and Discussion} \label{sec3}

\begin{figure*}[htb]
\centering
\includegraphics[width=\textwidth]{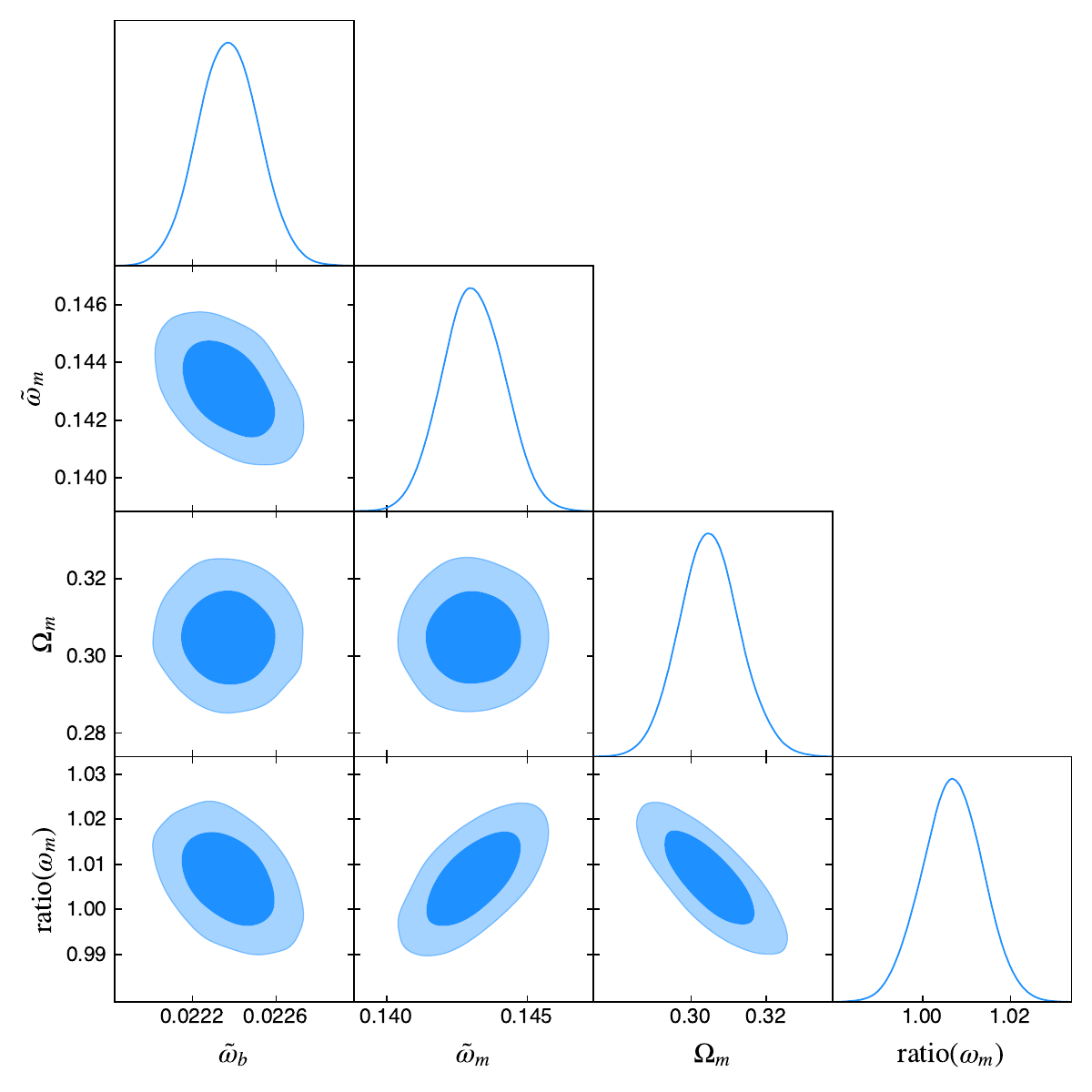}
\captionsetup{justification=raggedright}
\caption{Posterior likelihoods (68\% and 95\% confidence level) for the final parameter set considering the full samples of CMB and BAO.}\label{fig1}
\end{figure*}

The limitations of the parameters using BAO+CMB are illustrated in FIG.\ref{fig1}, and the results are also summarized in TABLE \ref{tab:table2}. Notably, the parameters of initial set show complex degeneracy compared to those of the final set, particularly for early parameters such as $\tilde{\Omega}_b$, $\tilde{\Omega}_m$ and $\tilde{H}_0$. The fundamental reason for the degeneracy was elaborated in the previous section.

\begin{table} 
\captionsetup{justification=raggedright}
\begin{ruledtabular}
\begin{tabular}{cc}
 Parameters & Value \\ \hline
    $\tilde{\omega}_b$ & $0.02237\pm0.00014$ \\
    $\tilde{\omega}_m$ & $0.1430\pm0.0011$  \\
    $\Omega_m$ & $0.3049\pm0.0081$ \\
    ${\rm ratio}(\omega_m)$ & $1.0069\pm0.0070$ \\
\end{tabular}
\caption{\label{tab:table2} Cosmological constraints on early and late parameters simultaneously.}
\end{ruledtabular}
\end{table}

\subsection{Comparison with separate constraints}
There are early parameters and late parameters incorporated in our model, enabling the simultaneous constraint of both. Note that, there are substantial cosmological parameters estimations obtained from high-redshift or low-redshift data alone. Therefore, we firstly compare the results presented in this paper with other results which are constrained separately before discussing on the ${\rm ratio}(\omega_m)$ and cosmological models.

Initially, we concentrate on the physical density of the early universe, denoted by $\tilde{\omega}_m=0.1430\pm0.0011$. Another recent estimation of early cosmological parameters from CMB observations almost independently of the late-time evolution yields $\tilde{\omega}_m=0.1414\pm0.0013$ \cite{lemos2023cmbearly}. They accomplish it by constructing a CMB likelihood that does not require a cosmological model to model the late-time effects. This indicates that the estimation of early parameters obtained independently is consistent with the simultaneous constraint.

In contrast of degeneracy of early parameters, we can analyze late parameters $H_0$ and $\Omega_m$ individually. Their posteriors are derived from $\tilde{\omega}_m$, $\Omega_m$ and ${\rm ratio}(\omega_m)$ showed in TABLE \ref{tab:table2}, and given by
\begin{eqnarray} \label{eq-24}
     H_0 &=& 68.34 \pm 0.97  \rm \ km\ s^{-1} Mpc^{-1}, \nonumber \\
     \Omega_m &=& 0.3049 \pm 0.0081  .
\end{eqnarray}

We can compare them with these values constrained by low-redshift observations, such as Supernovae. At first, we concentrate on the matter density $\Omega_m$ and adopt $\Omega_m = 0.298 \pm 0.022$ from the Pantheon analysis \cite{scolnic2018pantheon}. It is apparent that the Pantheon result is in agreement with ours. We also notice that constraint of $\Omega_m$ in this work is also consistent with the result \cite{pogosian2020baowithprior} $\Omega_m = 0.294^{+0.014}_{-0.016}$ constrained by BAO with a prior of $\omega_m$. 

We divide the BAO measurements into early and late portions to limit the value of the Hubble constant, $H_0$, using information from the late universe. Therefore it is interesting that compare it with $H_0$ measurements which are derived from local observation of SNIa, such as the SH0ES Team's analysis. They measured via a three-step distance ladder employing a single, simultaneous fit between geometric distance measurements to standardized Cepheid variables, standardized Cepheids and colocated SNIa in nearby galaxies, and SNIa in the Hubble flow. Namely the analysis employs a direct and cosmological model-independent measurement, refers $H_0 = 73.04 \pm 1.04 \rm \ km\ s^{-1} Mpc^{-1}$. The $3.3\sigma$ tension with constraint in this paper is not strange, because the late Hubble constant $H_0$ is not only constrained by the low redshift information of the BAO, but also includes the help of the CMB. In fact, this tension is still the well-known Hubble tension. However, we have found that replacing Cepheid variables with TRGB method \cite{freedman2021trgb} to calibrate SNIa yields a consistent value of $H_0 = 69.8 \pm 1.7 \rm \ km\ s^{-1} Mpc^{-1}$. We expect the local measurements of $H_0$ will be more precise in the future.

% Galaxy results
\begin{figure*}[htb]
\centering
\includegraphics[width=\textwidth]{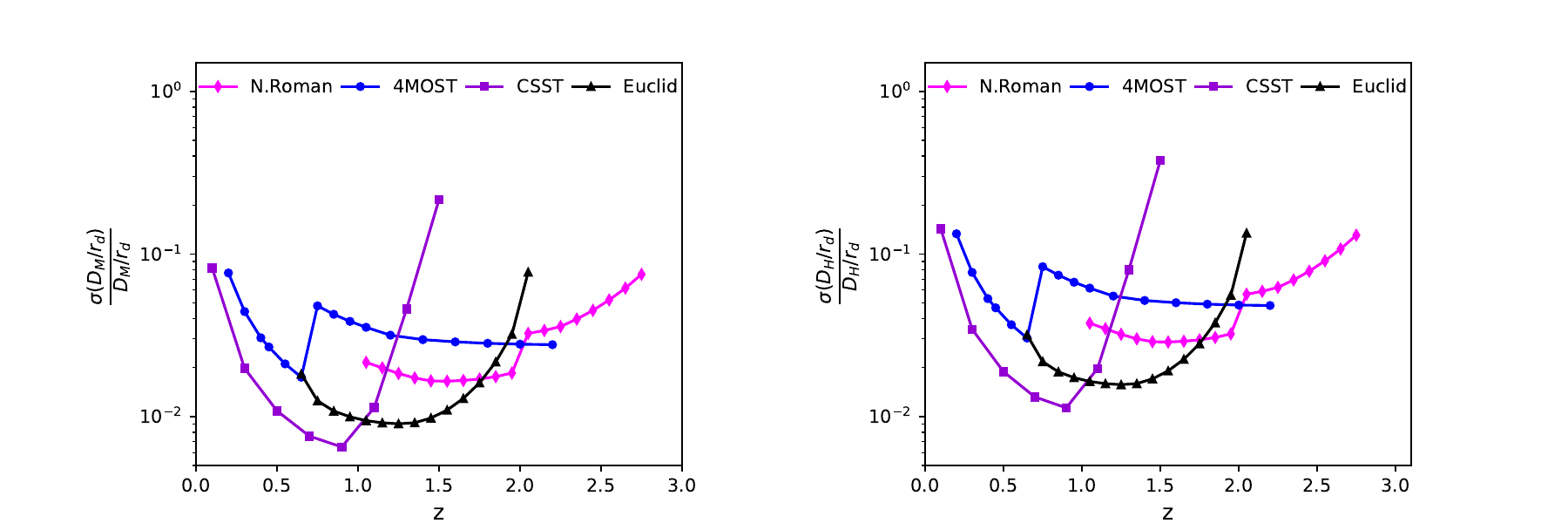}
\captionsetup{justification=raggedright}
\caption{Fractional errors on $D_M(z)/r_d$ and $D_H(z)/r_d$ for galaxy surveys, as a function of redshift.} \label{fig3}
\end{figure*}

%\begin{figure*}[htb]
%\centering
%\includegraphics[width=\textwidth]{BAO_simulation(Gal_comparison).pdf}
%\captionsetup{justification=raggedright}
%\caption{Comparison of fractional errors when using fitting formula Eq.\eqref{eq-25}. } %\label{fig4}
%\end{figure*}

\begin{figure}[htb]
\centering
\includegraphics[width=0.4\textwidth,height=0.282\textheight]{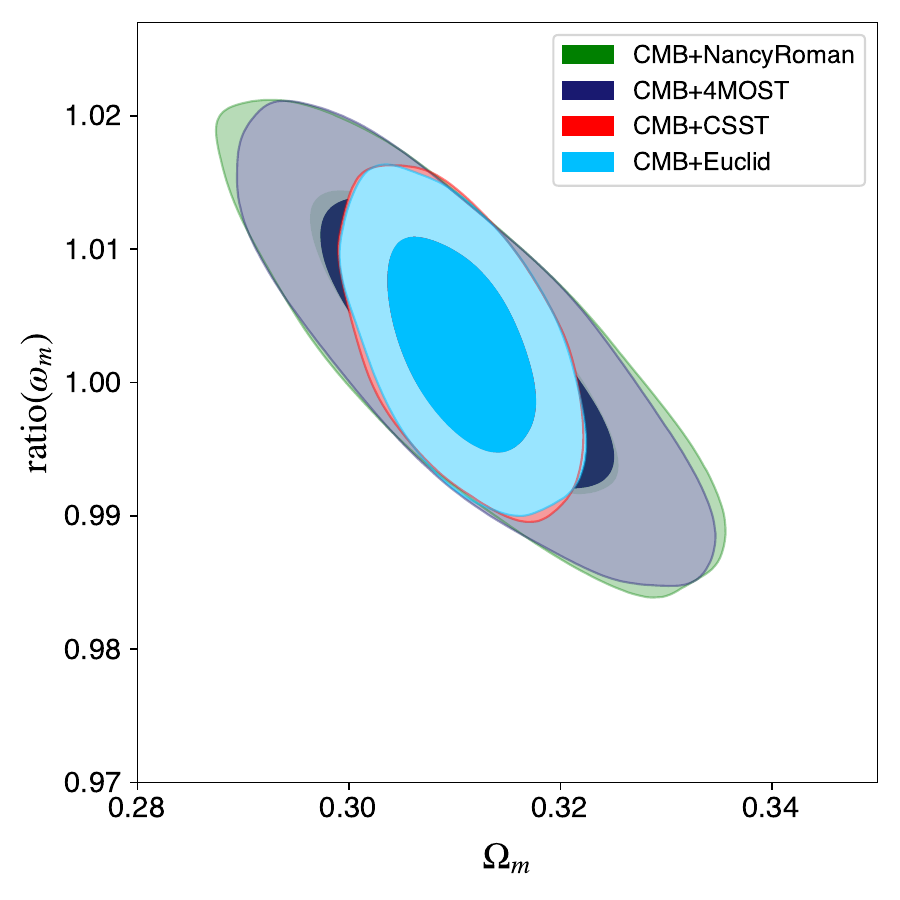}
\captionsetup{justification=raggedright}
\caption{Constraints (68\% and 95\% confidence level) on ${\rm ratio}(\omega_m)$ by using CMB+Nancy Roman, CMB+4MOST, CMB+CSST and CMB+Euclid.} \label{fig5}
\end{figure}

\subsection{Test of $\Lambda$CDM model}
% physical value
As mentioned above, differences in early or late cosmological parameters solely depend on the redshift of the measurements used in the constraint. Namely the two values of a cosmological parameter should be identical, or else the expansion history described by the $\Lambda$CDM model is not compatible. We take $\omega_m$ as a representative to analyze the potential difference between the early and late universe due to parameter degeneracy. We find that 
\begin{equation}
{\rm ratio}(\omega_m) = 1.0069\pm0.0070,
\end{equation}
with the unit falling within the $1\sigma$ range. This result indicates there is no discrepancy between the early and late epochs of the universe in $\Lambda$CDM model. In other words, $\omega_m$ have no redshift-dependence as $\Lambda$CDM requires. It means our test does not expose the defects of the $\Lambda$CDM model. %Although the Hubble tension still exists, the $\Lambda$CDM model is presently the most suitable form for describing the Universe.

\begin{figure*}[htb]
\centering
\includegraphics[width=\textwidth]{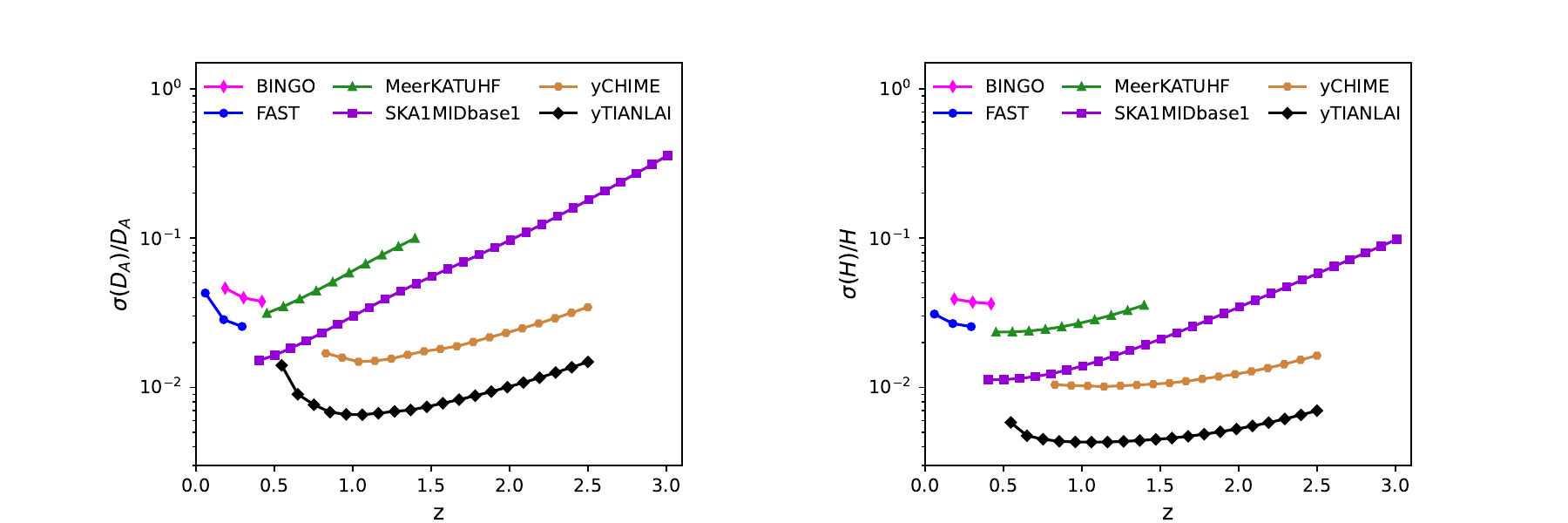}
\captionsetup{justification=raggedright}
\caption{Fractional errors on $D_A(z)$ and $H(z)$ for 21 cm IM surveys, as a function of redshift.} \label{fig6}
\end{figure*}

\begin{figure}[htb]
\centering
\includegraphics[width=0.4\textwidth,height=0.282\textheight]{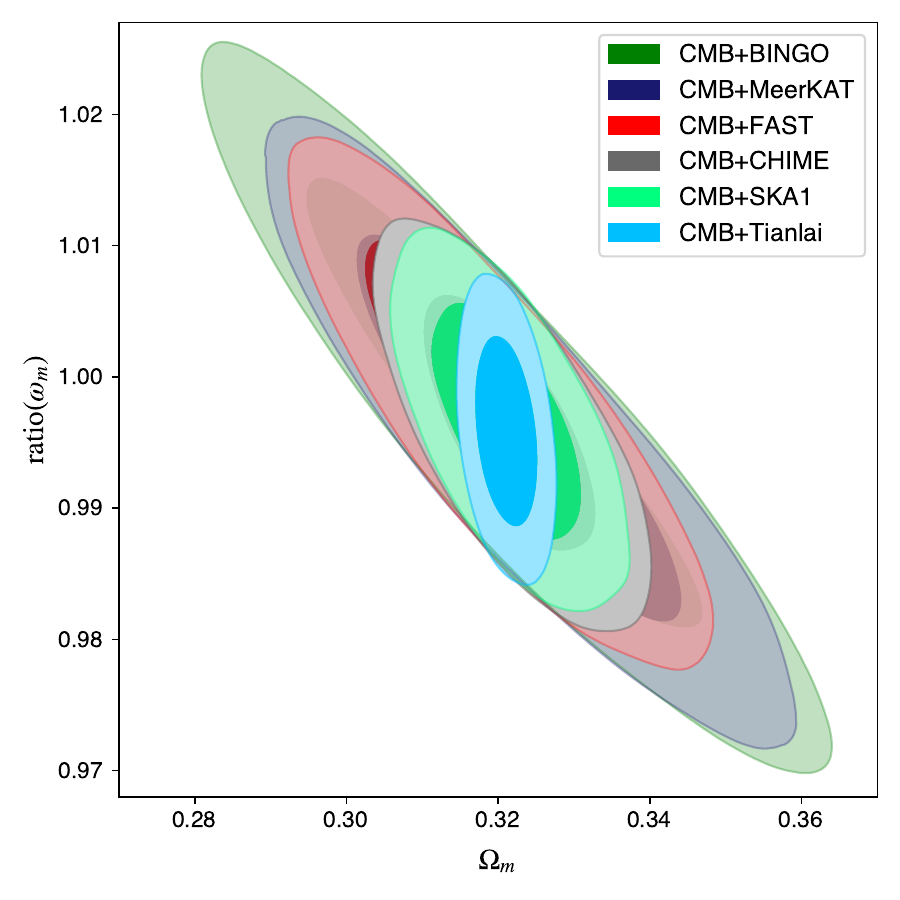}
\captionsetup{justification=raggedright}
\caption{Constraints (68\% and 95\% confidence level) on ${\rm ratio}(\omega_m)$ by using CMB+BINGO, CMB+MeerKAT, CMB+FAST, CMB+CHIME, CMB+SKA and CMB+Tianlai.} \label{fig7}
\end{figure}

\section{Forecasts for the future surveys} \label{sec4}
We aim to analyze the surveys that measures BAO features with higher accuracy, assessing their potential to enhance precison in our cosmological parameter estimates ${\rm ratio}(\omega_m)$. In this section, we consider two kinds of potention observations on BAO: the galaxy surveys and the neutral hydrogen surveys. 

The method with galaxy surveys has been discussed above, which catalog individual galaxies in angle and redshift to construct a map of matter distribution. While the method with neutral hydrogen surveys is proposed in \cite{chang2008beginofim}, which detects structures in the universe traced by the redshifted diffuse 21 cm hyperfine transition line of neutral hydrogen ($\rm H_I$) using the intensity mapping (IM) technique. The idea is to measure the $\rm H_I$ brightness temperature field to obtain information on the high levels of hydrogen abundance. This will result in a hydrogen map that can serve as a distinct tracer from galaxies, aiding in the measurement of the BAO scale. In both cases, the fiducial $\Lambda$CDM flat cosmological model with parameters is inferred by \textit{Planck} \cite{2020planck} is used in the analysis, including $\omega_b = 0.02242$, $\Omega_m = 0.3111$, $H_0 = 67.66 \rm \ km\ s^{-1} Mpc^{-1}$, $r_d = 147.09\ \rm Mpc$.

\subsection{Galaxy Surveys} % method for produce simulation; 
We use the analytic fitting formula developed by \cite{blake2006formulae} to depict the standard ruler accuracies in terms of the galaxy redshift survey parameters. Their analysis is based on Monte Carlo realizations \cite{blake2003bg03,glazebrook2005detailsforbg03} and the covariance matrix evaluated in \cite{peacock1994reconstructing}. Three configuring parameters are required for spectroscopic redshift surveys: central redshift $z$, total volume $V$ (in $h^{-3}\rm Gpc^3$), average number density of galaxies $n$ (in $10^{-3}h^3\rm Mpc^{-3}$). The formula is 
\begin{equation} \label{eq-25}
     x = x_0\sqrt{\frac{V_0}{V}} \left( 1+\frac{n_{eff}}{n}\frac{D(z_0)^2}{b(z)^2D(z)^2} \right),
\end{equation}
where $x$ is the fractional error, i.e. if $A$ is a measurement thus $x\equiv\sigma_A/A$. $x_0$ is a normalized coefficient and fiducial survey volume $V_0=2.16 h^{-3}\rm Gpc^3$. $D(z_0)=0.61$ is linear growth factor defined in Ref.\cite{carroll1992Dz} and $b(z)D(z)$ is bias factor which describe the difference between galaxy distribution with dark matter distribution. Note that, the fitting formula needs to reflect the effect of the non-linear evolution via an empirical power-law in redshift $z$
\begin{eqnarray}
     x &=& x_0\sqrt{\frac{V_0}{V}} \left( 1+\frac{n_{eff}}{n}\frac{D(z_0)^2}{b(z)^2D(z)^2} \right)\left(\frac{z_m}{z}\right)^\gamma z<z_m \nonumber \\ 
       &=& x_0\sqrt{\frac{V_0}{V}} \left( 1+\frac{n_{eff}}{n}\frac{D(z_0)^2}{b(z)^2D(z)^2} \right)  z>z_m
       \label{DETF}
\end{eqnarray}
where $z_m$ is the cut-off redshift and the effective number density of galaxies $n_{eff}$ is
\begin{eqnarray}
     n_{eff} &=& n_0\left[1-b\left(1-\frac{z}{z_m} \right)\right] \quad z<z_m
 \nonumber \\ 
             &=& n_0 \qquad \qquad \qquad \qquad \quad \ z>z_m
       \label{neff}
\end{eqnarray}

The fitting formula contains five parameters $\left(x_0, n_0, b, \gamma, z_m \right)$, which are listed in Table \ref{tab:detf}.

\begin{table} 
\captionsetup{justification=raggedright}
\begin{ruledtabular}
\begin{tabular}{c|c|c}
 Parameter & Spec-z Tangential  & Spec-z Radial   \\ \hline
    $x_0$  & $0.0085$           & $0.0148$   \\
    $n_0(\times10^{-3}h^3\rm Mpc^{-3})$  & $0.82$ & $0.82$ \\
    $z_m$ & $1.4$ & $1.4$    \\
    $\gamma$ & $0.5$ & $0.5$ \\
    $b$ & $0.52$ & $0.52$
\end{tabular}
\caption{\label{tab:detf} Best-fitting coefficients for the fitting formula from Ref.\cite{blake2006formulae}, for the two types of standard ruler accuracy:$D_H(z)/r_d$ and $D_M(z)/r_d$. }
\end{ruledtabular}
\end{table}

Four surveys with distinct objectives are considered, including 4MOST \footnote{https://4MOST} (4-meter Multi-Object Spectroscopic Telescope) \cite{richard20194most}, Roman \footnote{http://roman.gsfc.nasa.gov} (Nancy Grace Roman Space Telescope) \cite{spergel2013wfirst}, CSST \footnote{http://www.nao.cas.cn/csst/} (The China Space Station Telescope) \cite{zhan2011csst1,zhan2021csst2} and Euclid \footnote{https://www.euclid-ec.org} \cite{laureijs2011euclid1,blanchard2020euclid2}. The configuration parameters are listed in Table \ref{tab:table3}. 

\begin{table}[h]
\captionsetup{justification=raggedright}
\begin{ruledtabular}
\begin{tabular}{ccccc}
         & 4MOST & N.Roman & CSST & Euclid \\ \hline
$z_{\rm min}$ & 0.15 & 1.0 & 0 & 0.6 \\
$z_{\rm max}$ & 2.2 & 2.8 & 1.6 & 2.1 \\
$S_{\rm area}[\rm deg^2]$ & 7500(1000) & 2000 & 17500 & 15000 \\
$n$ & \cite{richard20194most} & \cite{font2014desi} & \cite{ding2023csstdata} & \cite{font2014desi} \\
\end{tabular}
\caption{\label{tab:table3}The configurations of spectroscopic galaxy surveys used in this paper. $n$ is the number density of galaxies. Note that 4MOST observe different sources in different sky coverage area.}
\end{ruledtabular}
\end{table}

%  simulation data analysis
For N.Roman and Euclid we use the redshift bin $\Delta z = 0.1$ and for CSST $\Delta z = 0.2$, 4MOST use both for different targets. The bias factor $b(z)D(z)$ is also from the same reference listed in TABLE \ref{tab:table3}, including $b(z)D(z)=0.76$ for N.Roman and Euclid and a linear galaxy bias as $1+0.84z$ for CSST, a set $(1.34, 1.7, 0.84, 1.2)$ for different targets of 4MOST. And the mean value of BAO measurements can be calculated based on the fiducial model. Then, by using the fitting formula Eq.\eqref{DETF}, we can obtain the fractional errors for both transverse measurement $D_M(z)/r_d$ and line-of-sight measurement $D_H(z)/r_d$. FIG \ref{fig3} shows the features of different surveys: 4MOST covers a large redshift range and CSST performs optimally for observing low-redshift objects $(z<1)$. However, in redshift range $1<z<2$ Euclid has advantages and Roman are precise enough in the measurement of high redshift galaxies $(z>2)$. %In contrast to simulation results based on the complete Fisher matrix method (using power spectrum) \cite{seo2007baoforecast} \cite{font2014desi}, FIG \ref{fig4} illustrates the trends of fractional error are similar to each other. The errors derived from the fitting formula are slightly larger than those determined by the complete Fisher matrix method \cite{font2014desi}. ??????? which result is used?????

% constrained results analysis 
We perform a Markov Chain Monte Carlo analysis using our simulation data generated with the CMB distance prior to constrain cosmological parameters. The forecast results using different surveys are plotted in FIG \ref{fig5} and list $1\sigma$ errors of ${\rm ratio}(\omega_m)$ in TABLE \ref{tab:table4}. We can find that Roman and 4MOST are similar in constraining the ratio of $\omega_m$. Thus, either Roman $\sigma({\rm ratio}(\omega_m))=0.0075$ or 4MOST $\sigma({\rm ratio}(\omega_m))=0.0072$ alone have comparable precision with the combination of current BAO probes $\sigma({\rm ratio}(\omega_m))=0.0070$. CSST and Euclid can tighten the constraints due to their large sky cover area. Euclid gives the best constraints in galaxy redshift surveys, $\sigma({\rm ratio}(\omega_m))=0.0050$, which means Euclid alone can achieve about 30 percent improvement in precision.

\begin{table}[h]
\captionsetup{justification=raggedright}
\begin{ruledtabular}
\begin{tabular}{cc}
 Surveys & $\sigma({\rm ratio}(\omega_m))$ \\ \hline
    Nancy Roman & 0.0075  \\
    4MOST & 0.0072 \\
    CSST & 0.0054 \\
    Euclid & 0.0050 \\
\end{tabular}
\caption{\label{tab:table4}Forecast $1\sigma$ errors of ${\rm ratio}(\omega_m)$ derived from future galaxy redshift surveys combined with \textit{Planck} CMB data.}
\end{ruledtabular}
\end{table}

\subsection{21 cm IM Surveys} % generating methods + simulation data analysis
We use the public code \texttt{RadioFisher} \cite{bull2015RadioFisher} to obtain the Fisher matrix for $D_A(z)$ and $H(z)$, allowing us to invert it and derive the covariance matrix for observations of future IM surveys. The core of code is the construction and calculation of the Fisher matrix. For a cosmological parameter set $\{p\}$, the Fisher matrix can be expressed as \cite{seo2007baoforecast,bull2015RadioFisher,wu2022prospects}
\begin{equation}\label{Fisher}
    F_{ij} = \frac{V_{\rm bin}}{8\pi^2}\int_{-1}^{1}\mathrm{d}\mu \int_{k_{min}}^{k_{max}}k^2\mathrm{d}k \frac{\partial \ln C^T}{\partial p_i}\frac{\partial \ln C^T}{\partial p_j},
\end{equation}
where $V_{\rm bin}$ is the physical volume, $\mu$ is the cosine of the angle between the line of sight and the direction of $k$. The parameter set $\{p\}$ of this analysis includes the angular diameter distance $D_A(z)$, Hubble parameter $H(z)$, redshift-space distortion (RSD) observable $[f\sigma_8](z)$, $\rm H_I$ bias parameter $[b_{HI}\sigma_8](z)$ and non-linear dispersion scale parameter $\sigma_{\rm NL}$. And $C^T$ is the total covariance of 21 cm emission line measurements, including the contribution of signal $C^S$, noise $C^N$ and foreground $C^F$. 

The signal model can be summarized simply as
\begin{eqnarray} \label{signal}
    C^S(z,k)&=&T_b^2(z)P_{tot}(k) \\ \nonumber
            &=&T_b^2(z)F_{RSD}(k)D^2(z)P(k,z=0),
\end{eqnarray}
where $T_b$ is the effective $\rm H_I$ brightness temperature and $P_{tot}(k)$ is the matter power spectrum with the effect of redshift space distortion (RSD). Note that, the signal covariance is totally a theoretical calculation which is determined by the cosmological parameters mentioned above. However, the noise covariance is based on the configuration of experiments
\begin{equation}\label{noise}
    C^N(k)=\sigma^2_{pix}(N_d,N_b,T_{inst},D_d)V_{pix}B(k),
\end{equation}
where $\sigma_{pix}$ and $V_{pix}$ is the pixel noise temperature and volume. The noise temperature $\sigma_{pix}$ depends on configuration parameters, including the receiver instrument noise temperature $T_{\rm inst}$, the number of dishes $N_d$ and beams $N_b$, diameter of the dish $D_d$. And volume $V_{pix}$ relates to the survey area $S_{\rm area}$.  The factor $B(k)$ is used to describe the frequency and angular responses of the instrument. The residual foreground covariance $C^F$ is controlled by a scaling factor $\varepsilon_{FG}$ which describes the efficiency of the foreground removal process. With the help of \texttt{RadioFisher}, we can calculate the total covariance $C^T=C^S+C^N+C^F$ and derive the uncertainties of BAO measurements from Fisher matrix. More details are discussed in Sec.2 of Ref.\cite{wu2022prospects}.

We determine several 21 cm IM surveys, including single-dish surveys BINGO \footnote{https://bingotelescope.org} (Baryon Acoustic Oscillations from Integrated Neutral Gas Observations) \cite{battye2013bingo1, wuensche2020bingo2, abdalla2022bingo3}, FAST \footnote{https://fast.bao.ac.cn} (Five-hundred-meter Aperture Spherical Telescope) \cite{nan2011fast1, smoot2017fast2} and dish interferometer surveys MeerKAT radio telescope \footnote{https://meerkat} \cite{jonas2009meerkat1, santos2015meerkat2, bacon2020meerkat3}, SKA \footnote{https://www.skatelescope.org} (Square Kilometre Array) \cite{dewdney2013ska} and cylinder interferometer surveys CHIME \footnote{https://chime-experiment.ca} (Canadian Hydrogen Intensity Mapping Experiment) \cite{newburgh2014chime1, bandura2014chime2}, Tianlai \footnote{http://tianlai.bao.ac.cn} \cite{chen2012tianlai1, li2020tianlai2}. The configuration parameters of 21 cm IM surveys are listed in Table \ref{tab:IM}. 

\begin{table}[h]
\captionsetup{justification=raggedright}
\begin{ruledtabular}
\begin{tabular}{ccccccc}
         & BINGO & FAST & MeerKAT & SKA & CHIME & Tianlai \\ \hline
$N_d$ & $1$ & $1$ & $64$ & $190$ & $1280$ & $2048$ \\
$N_b$ & $50$ & $20$ & $1$ & $1$ & $1$ & $1$ \\
$T_{\rm inst}\ [\rm K]$ & 50 & 20 & 29 & 28 & 50 & 50 \\
$D_{\rm dish}\ [\rm m]$ & 40 & 300 & 13.5 & 15 & 20 & 15 \\
$z_{\rm min}$ & 0.13 & 0 & 0.4 & 0.35 & 0.77 & 0.49\\
$z_{\rm max}$ & 0.45 & 0.35 & 1.45 & 3 & 2.55 & 2.55\\
$S_{\rm area}[\rm deg^2]$ & 3000 & 20000 & 25000 & 25000 & 20000 & 20000 \\
\end{tabular}
\caption{\label{tab:IM}The configurations of 21 cm IM surveys used in this paper is mainly from Ref.\cite{bull2015RadioFisher}.}
\end{ruledtabular}
\end{table}

% explain the reason of difference
%Although the same configuration parameters are used as \cite{wu2022prospects}, we choose a different parameter set and contain $\lbrace D_A(z), H(z), [f\sigma_8](z), [b_{HI}\sigma_8](z), \sigma_{NL} \rbrace$ after marginalization. Therefore our forecast results for $D_A(z)$ and $H(z)$ showed in FIG \ref{fig6} differ from those presented in Figure 2 of \cite{wu2022prospects}.

%footnote: SKA is SKA1-MID 
% constrained results analysis
Because the rest frame frequency of the 21 cm line is $\nu_{21}=1420.4 MHz$, the redshift range is often transformed to the range of frequency. The bin width is chosen as $\Delta \nu=60MHz$ ($\Delta z\approx0.1$) for all surveys and the same fiducial model is used to calculate the mean value of $D_A(z)$ and $H(z)$. The fractional errors of the BAO measurements are showed in FIG.\ \ref{fig6} and it is clear Tianlai have the most precise measurements. We perform a Markov Chain Monte Carlo analysis using our simulation data generated with the CMB distance prior to constrain cosmological parameters. From the contours displayed in FIG. \ref{fig7} and $1\sigma$ errors summarized in TABLE \ref{tab:table5}, BINGO and FAST provide weak constraints on ${\rm ratio}(\omega_m)$ due to their narrow redshift range. Additionally, the constraint given by MeerKAT is between them because it covers a small sky area in contrast to FAST or SKA. However, CHIME, SKA and Tianlai performs tighter constraint than the combination of current BAO probes, with $\sigma({\rm ratio}(\omega_m))=0.0063,\ 0.0058,\ 0.0047$ respectively, for their wide redshift ranges and high spatial resolution. Especially Tianlai enhances the ability of constraining ${\rm ratio}(\omega_m)$ significantly, which means we can investigate the potential discrepancy between early and late universe more accurately.

\begin{table}[h]
\captionsetup{justification=raggedright}
\begin{ruledtabular}
\begin{tabular}{cc}
 Surveys & $\sigma({\rm ratio}(\omega_m))$ \\ \hline
    BINGO & 0.0110  \\
    MeerKAT & 0.0098 \\
    FAST & 0.0082 \\
    CHIME & 0.0063 \\
    SKA & 0.0058 \\
    Tianlai & 0.0047 \\
\end{tabular}
\caption{\label{tab:table5}Forecast $1\sigma$ errors of ${\rm ratio}(\omega_m)$ derived from future 21 cm IM surveys combined with \textit{Planck} CMB data.}
\end{ruledtabular}
\end{table}

\section{Conclusions} \label{sec5}  % origin + specific steps; results analysis; forecast 
In recent years, the Hubble tension between the indirect model dependent estimates at early time and the direct late time model-independent measurements have increased to $5\sigma$. This huge discrepency prompt us to investigate its origin. Therefore, we desire to know whether the standard cosmological model is flawed or not through the consistency test of the cosmological parameter $\omega_m$. If $\omega_m$ changes with redshift, then we can conclude $\Lambda$CDM model breaks down. We utilize two discrete sets of parameters describing the expansion history of the Universe in its early and late stages respectively. The recombination redshift $z_*$ is chosen as the dividing line to distinguish between the early and late epochs. %Accordingly, early parameters are constrained by measurements that satisfy $z>z_*$, such as sound horizon $r_s(z_*)$, while late parameters are constrained by measurements that satisfy $z<z_*$, such as BAO direct measurements $D_M(z_*)$ or $D_H(z)$. 
Because of the degeneracy, we convert the complete parameter set from $\lbrace \tilde{\Omega}_b,\tilde{\Omega}_m,\tilde{H}_0,\Omega_m,H_0 \rbrace$ to $\lbrace \tilde{\omega}_b,\tilde{\omega}_m,\Omega_m,{\rm ratio}(\omega_m) \rbrace$. 

The combination of BAO and CMB data gives constraints on both early and late parameters at the same time, allowing for a comparison with early and late parameters that are constrained separately. We find that our constraints are almost all consistent with separate constraints except $H_0$ derived from SH0ES. % The smaller error of $\Omega_m$ indicates that we improve the precision by distinguishing early and late cosmological parameters. 
The most important result is that the ratio of $\tilde{\omega}_m$ to $\omega_m$ can help us judge the consistency between early and late cosmological parameters. %Either $\tilde{\omega}_m$ or $\omega_m$ is the current value of $\omega_m$, which results in ${\rm ratio}(\omega_m)=1$. 
The result ${\rm ratio}(\omega_m)=1.0069 \pm 0.0070$ implicates the early epoch and late epoch of the Universe described by $\Lambda$CDM are consistent. It means that  $\Lambda$CDM expose no defects in our test and there is no evidence for new physics beyond $\Lambda$CDM model. %We find ${\rm ratio}(\omega_m)$ should be $1.075\pm0.056$ if the Universe is described by EDE model. Therefore, the $1.2\sigma$ disagreement is difficult to determine the existence of early dark energy in the early universe.

To assess the level of precision to which ${\rm ratio}(\omega_m)$ can be constrained, simulation data of future BAO observations has been generated. This includes galaxy redshift surveys as well as 21 cm Intensity Mapping (IM) surveys. %We have used the fitting formula \cite{blake2006formulae} for galaxy surveys and codes developed by \cite{bull2015RadioFisher} for 21 cm IM surveys. 
In galaxy surveys, we find Euclid gives the best constraint with a $30\%$ improvement in precision, $\sigma({\rm ratio}(\omega_m))=0.0050$. In 21 cm IM surveys, SKA and Tianlai significantly tighten the constraints with $\sigma({\rm ratio}(\omega_m))=0.0058$ and $\sigma({\rm ratio}(\omega_m))=0.0047$. Taking into account new future CMB observations, the consistency of cosmological parameters will face a challenging assessment. We hope that these powerful future observations can shed light on the potential new physic beyond $\Lambda$CDM cosmology and the origin of the Hubble tension. 

~

\begin{acknowledgments}
We appreciate the helpful discussion with Philip Bull, Salvatore Capozziello and Eoin O Colgain. This work is supported by the National Key R\&D Program of China (Grant No. 2021YFC2203102 and 2022YFC2204602), Strategic Priority Research Program of the Chinese Academy of Science (Grant No. XDB0550300), the National Natural Science Foundation of China (Grant No. 12325301 and 12273035), the Fundamental Research Funds for the Central Universities (Grant No. WK2030000036 and WK3440000004), the Science Research Grants from the China Manned Space Project (Grant No.CMS-CSST-2021-B01), the 111 Project for "Observational and Theoretical Research on Dark Matter and Dark Energy" (Grant No. B23042).
\end{acknowledgments}
% 2 colors/ratio of figure; texttt 
% radio survey redshift ： 21cm is special 

% DH/rd 为例解释BAO简并关系；2层简并CMB可解一层 completed
% CMB部分 只解释l_a即可；althogh.... so our final set is only...
% new plots

\bibliography{referclean}% Produces the bibliography via BibTeX.

\begin{thebibliography}{75}%
\makeatletter
\providecommand \@ifxundefined [1]{%
 \@ifx{#1\undefined}
}%
\providecommand \@ifnum [1]{%
 \ifnum #1\expandafter \@firstoftwo
 \else \expandafter \@secondoftwo
 \fi
}%
\providecommand \@ifx [1]{%
 \ifx #1\expandafter \@firstoftwo
 \else \expandafter \@secondoftwo
 \fi
}%
\providecommand \natexlab [1]{#1}%
\providecommand \enquote  [1]{``#1''}%
\providecommand \bibnamefont  [1]{#1}%
\providecommand \bibfnamefont [1]{#1}%
\providecommand \citenamefont [1]{#1}%
\providecommand \href@noop [0]{\@secondoftwo}%
\providecommand \href [0]{\begingroup \@sanitize@url \@href}%
\providecommand \@href[1]{\@@startlink{#1}\@@href}%
\providecommand \@@href[1]{\endgroup#1\@@endlink}%
\providecommand \@sanitize@url [0]{\catcode `\\12\catcode `\$12\catcode `\&12\catcode `\#12\catcode `\^12\catcode `\_12\catcode `\%12\relax}%
\providecommand \@@startlink[1]{}%
\providecommand \@@endlink[0]{}%
\providecommand \url  [0]{\begingroup\@sanitize@url \@url }%
\providecommand \@url [1]{\endgroup\@href {#1}{\urlprefix }}%
\providecommand \urlprefix  [0]{URL }%
\providecommand \Eprint [0]{\href }%
\providecommand \doibase [0]{http://dx.doi.org/}%
\providecommand \selectlanguage [0]{\@gobble}%
\providecommand \bibinfo  [0]{\@secondoftwo}%
\providecommand \bibfield  [0]{\@secondoftwo}%
\providecommand \translation [1]{[#1]}%
\providecommand \BibitemOpen [0]{}%
\providecommand \bibitemStop [0]{}%
\providecommand \bibitemNoStop [0]{.\EOS\space}%
\providecommand \EOS [0]{\spacefactor3000\relax}%
\providecommand \BibitemShut  [1]{\csname bibitem#1\endcsname}%
\let\auto@bib@innerbib\@empty
%</preamble>
\bibitem [{\citenamefont {Aghanim}\ \emph {et~al.}(2020)\citenamefont {Aghanim}, \citenamefont {Akrami}, \citenamefont {Ashdown}, \citenamefont {Aumont}, \citenamefont {Baccigalupi}, \citenamefont {Ballardini}, \citenamefont {Banday}, \citenamefont {Barreiro}, \citenamefont {Bartolo}, \citenamefont {Basak} \emph {et~al.}}]{2020planck}%
  \BibitemOpen
  \bibfield  {author} {\bibinfo {author} {\bibfnamefont {N.}~\bibnamefont {Aghanim}}, \bibinfo {author} {\bibfnamefont {Y.}~\bibnamefont {Akrami}}, \bibinfo {author} {\bibfnamefont {M.}~\bibnamefont {Ashdown}}, \bibinfo {author} {\bibfnamefont {J.}~\bibnamefont {Aumont}}, \bibinfo {author} {\bibfnamefont {C.}~\bibnamefont {Baccigalupi}}, \bibinfo {author} {\bibfnamefont {M.}~\bibnamefont {Ballardini}}, \bibinfo {author} {\bibfnamefont {A.}~\bibnamefont {Banday}}, \bibinfo {author} {\bibfnamefont {R.}~\bibnamefont {Barreiro}}, \bibinfo {author} {\bibfnamefont {N.}~\bibnamefont {Bartolo}}, \bibinfo {author} {\bibfnamefont {S.}~\bibnamefont {Basak}},  \emph {et~al.},\ }\href@noop {} {\bibfield  {journal} {\bibinfo  {journal} {Astronomy \& Astrophysics}\ }\textbf {\bibinfo {volume} {641}},\ \bibinfo {pages} {A6} (\bibinfo {year} {2020})}\BibitemShut {NoStop}%
\bibitem [{\citenamefont {Balkenhol}\ \emph {et~al.}(2021)\citenamefont {Balkenhol}, \citenamefont {Dutcher}, \citenamefont {Ade}, \citenamefont {Ahmed}, \citenamefont {Anderes}, \citenamefont {Anderson}, \citenamefont {Archipley}, \citenamefont {Avva}, \citenamefont {Aylor}, \citenamefont {Barry} \emph {et~al.}}]{balkenholspt}%
  \BibitemOpen
  \bibfield  {author} {\bibinfo {author} {\bibfnamefont {L.}~\bibnamefont {Balkenhol}}, \bibinfo {author} {\bibfnamefont {D.}~\bibnamefont {Dutcher}}, \bibinfo {author} {\bibfnamefont {P.}~\bibnamefont {Ade}}, \bibinfo {author} {\bibfnamefont {Z.}~\bibnamefont {Ahmed}}, \bibinfo {author} {\bibfnamefont {E.}~\bibnamefont {Anderes}}, \bibinfo {author} {\bibfnamefont {A.}~\bibnamefont {Anderson}}, \bibinfo {author} {\bibfnamefont {M.}~\bibnamefont {Archipley}}, \bibinfo {author} {\bibfnamefont {J.}~\bibnamefont {Avva}}, \bibinfo {author} {\bibfnamefont {K.}~\bibnamefont {Aylor}}, \bibinfo {author} {\bibfnamefont {P.}~\bibnamefont {Barry}},  \emph {et~al.},\ }\href@noop {} {\bibfield  {journal} {\bibinfo  {journal} {Physical Review D}\ }\textbf {\bibinfo {volume} {104}},\ \bibinfo {pages} {083509} (\bibinfo {year} {2021})}\BibitemShut {NoStop}%
\bibitem [{\citenamefont {Aiola}\ \emph {et~al.}(2020)\citenamefont {Aiola}, \citenamefont {Calabrese}, \citenamefont {Maurin}, \citenamefont {Naess}, \citenamefont {Schmitt}, \citenamefont {Abitbol}, \citenamefont {Addison}, \citenamefont {Ade}, \citenamefont {Alonso}, \citenamefont {Amiri} \emph {et~al.}}]{aiola2020act}%
  \BibitemOpen
  \bibfield  {author} {\bibinfo {author} {\bibfnamefont {S.}~\bibnamefont {Aiola}}, \bibinfo {author} {\bibfnamefont {E.}~\bibnamefont {Calabrese}}, \bibinfo {author} {\bibfnamefont {L.}~\bibnamefont {Maurin}}, \bibinfo {author} {\bibfnamefont {S.}~\bibnamefont {Naess}}, \bibinfo {author} {\bibfnamefont {B.~L.}\ \bibnamefont {Schmitt}}, \bibinfo {author} {\bibfnamefont {M.~H.}\ \bibnamefont {Abitbol}}, \bibinfo {author} {\bibfnamefont {G.~E.}\ \bibnamefont {Addison}}, \bibinfo {author} {\bibfnamefont {P.~A.}\ \bibnamefont {Ade}}, \bibinfo {author} {\bibfnamefont {D.}~\bibnamefont {Alonso}}, \bibinfo {author} {\bibfnamefont {M.}~\bibnamefont {Amiri}},  \emph {et~al.},\ }\href@noop {} {\bibfield  {journal} {\bibinfo  {journal} {Journal of Cosmology and Astroparticle Physics}\ }\textbf {\bibinfo {volume} {2020}},\ \bibinfo {pages} {047} (\bibinfo {year} {2020})}\BibitemShut {NoStop}%
\bibitem [{\citenamefont {Scolnic}\ \emph {et~al.}(2018)\citenamefont {Scolnic}, \citenamefont {Jones}, \citenamefont {Rest}, \citenamefont {Pan}, \citenamefont {Chornock}, \citenamefont {Foley}, \citenamefont {Huber}, \citenamefont {Kessler}, \citenamefont {Narayan}, \citenamefont {Riess} \emph {et~al.}}]{scolnic2018pantheon}%
  \BibitemOpen
  \bibfield  {author} {\bibinfo {author} {\bibfnamefont {D.~M.}\ \bibnamefont {Scolnic}}, \bibinfo {author} {\bibfnamefont {D.}~\bibnamefont {Jones}}, \bibinfo {author} {\bibfnamefont {A.}~\bibnamefont {Rest}}, \bibinfo {author} {\bibfnamefont {Y.}~\bibnamefont {Pan}}, \bibinfo {author} {\bibfnamefont {R.}~\bibnamefont {Chornock}}, \bibinfo {author} {\bibfnamefont {R.}~\bibnamefont {Foley}}, \bibinfo {author} {\bibfnamefont {M.}~\bibnamefont {Huber}}, \bibinfo {author} {\bibfnamefont {R.}~\bibnamefont {Kessler}}, \bibinfo {author} {\bibfnamefont {G.}~\bibnamefont {Narayan}}, \bibinfo {author} {\bibfnamefont {A.}~\bibnamefont {Riess}},  \emph {et~al.},\ }\href@noop {} {\bibfield  {journal} {\bibinfo  {journal} {The Astrophysical Journal}\ }\textbf {\bibinfo {volume} {859}},\ \bibinfo {pages} {101} (\bibinfo {year} {2018})}\BibitemShut {NoStop}%
\bibitem [{\citenamefont {Brout}\ \emph {et~al.}(2022)\citenamefont {Brout}, \citenamefont {Scolnic}, \citenamefont {Popovic}, \citenamefont {Riess}, \citenamefont {Carr}, \citenamefont {Zuntz}, \citenamefont {Kessler}, \citenamefont {Davis}, \citenamefont {Hinton}, \citenamefont {Jones} \emph {et~al.}}]{brout2022pantheon+}%
  \BibitemOpen
  \bibfield  {author} {\bibinfo {author} {\bibfnamefont {D.}~\bibnamefont {Brout}}, \bibinfo {author} {\bibfnamefont {D.}~\bibnamefont {Scolnic}}, \bibinfo {author} {\bibfnamefont {B.}~\bibnamefont {Popovic}}, \bibinfo {author} {\bibfnamefont {A.~G.}\ \bibnamefont {Riess}}, \bibinfo {author} {\bibfnamefont {A.}~\bibnamefont {Carr}}, \bibinfo {author} {\bibfnamefont {J.}~\bibnamefont {Zuntz}}, \bibinfo {author} {\bibfnamefont {R.}~\bibnamefont {Kessler}}, \bibinfo {author} {\bibfnamefont {T.~M.}\ \bibnamefont {Davis}}, \bibinfo {author} {\bibfnamefont {S.}~\bibnamefont {Hinton}}, \bibinfo {author} {\bibfnamefont {D.}~\bibnamefont {Jones}},  \emph {et~al.},\ }\href@noop {} {\bibfield  {journal} {\bibinfo  {journal} {The Astrophysical Journal}\ }\textbf {\bibinfo {volume} {938}},\ \bibinfo {pages} {110} (\bibinfo {year} {2022})}\BibitemShut {NoStop}%
\bibitem [{\citenamefont {Alam}\ \emph {et~al.}(2017)\citenamefont {Alam}, \citenamefont {Ata}, \citenamefont {Bailey}, \citenamefont {Beutler}, \citenamefont {Bizyaev}, \citenamefont {Blazek}, \citenamefont {Bolton}, \citenamefont {Brownstein}, \citenamefont {Burden}, \citenamefont {Chuang} \emph {et~al.}}]{alam2017sdss3}%
  \BibitemOpen
  \bibfield  {author} {\bibinfo {author} {\bibfnamefont {S.}~\bibnamefont {Alam}}, \bibinfo {author} {\bibfnamefont {M.}~\bibnamefont {Ata}}, \bibinfo {author} {\bibfnamefont {S.}~\bibnamefont {Bailey}}, \bibinfo {author} {\bibfnamefont {F.}~\bibnamefont {Beutler}}, \bibinfo {author} {\bibfnamefont {D.}~\bibnamefont {Bizyaev}}, \bibinfo {author} {\bibfnamefont {J.~A.}\ \bibnamefont {Blazek}}, \bibinfo {author} {\bibfnamefont {A.~S.}\ \bibnamefont {Bolton}}, \bibinfo {author} {\bibfnamefont {J.~R.}\ \bibnamefont {Brownstein}}, \bibinfo {author} {\bibfnamefont {A.}~\bibnamefont {Burden}}, \bibinfo {author} {\bibfnamefont {C.-H.}\ \bibnamefont {Chuang}},  \emph {et~al.},\ }\href@noop {} {\bibfield  {journal} {\bibinfo  {journal} {Monthly Notices of the Royal Astronomical Society}\ }\textbf {\bibinfo {volume} {470}},\ \bibinfo {pages} {2617} (\bibinfo {year} {2017})}\BibitemShut {NoStop}%
\bibitem [{\citenamefont {Alam}\ \emph {et~al.}(2021)\citenamefont {Alam}, \citenamefont {Aubert}, \citenamefont {Avila}, \citenamefont {Balland}, \citenamefont {Bautista}, \citenamefont {Bershady}, \citenamefont {Bizyaev}, \citenamefont {Blanton}, \citenamefont {Bolton}, \citenamefont {Bovy} \emph {et~al.}}]{alam2021completed}%
  \BibitemOpen
  \bibfield  {author} {\bibinfo {author} {\bibfnamefont {S.}~\bibnamefont {Alam}}, \bibinfo {author} {\bibfnamefont {M.}~\bibnamefont {Aubert}}, \bibinfo {author} {\bibfnamefont {S.}~\bibnamefont {Avila}}, \bibinfo {author} {\bibfnamefont {C.}~\bibnamefont {Balland}}, \bibinfo {author} {\bibfnamefont {J.~E.}\ \bibnamefont {Bautista}}, \bibinfo {author} {\bibfnamefont {M.~A.}\ \bibnamefont {Bershady}}, \bibinfo {author} {\bibfnamefont {D.}~\bibnamefont {Bizyaev}}, \bibinfo {author} {\bibfnamefont {M.~R.}\ \bibnamefont {Blanton}}, \bibinfo {author} {\bibfnamefont {A.~S.}\ \bibnamefont {Bolton}}, \bibinfo {author} {\bibfnamefont {J.}~\bibnamefont {Bovy}},  \emph {et~al.},\ }\href@noop {} {\bibfield  {journal} {\bibinfo  {journal} {Physical Review D}\ }\textbf {\bibinfo {volume} {103}},\ \bibinfo {pages} {083533} (\bibinfo {year} {2021})}\BibitemShut {NoStop}%
\bibitem [{\citenamefont {Abbott}\ \emph {et~al.}(2018)\citenamefont {Abbott}, \citenamefont {Abdalla}, \citenamefont {Annis}, \citenamefont {Bechtol}, \citenamefont {Blazek}, \citenamefont {Benson}, \citenamefont {Bernstein}, \citenamefont {Bernstein}, \citenamefont {Bertin}, \citenamefont {Brooks} \emph {et~al.}}]{abbott2018desy1}%
  \BibitemOpen
  \bibfield  {author} {\bibinfo {author} {\bibfnamefont {T.}~\bibnamefont {Abbott}}, \bibinfo {author} {\bibfnamefont {F.}~\bibnamefont {Abdalla}}, \bibinfo {author} {\bibfnamefont {J.}~\bibnamefont {Annis}}, \bibinfo {author} {\bibfnamefont {K.}~\bibnamefont {Bechtol}}, \bibinfo {author} {\bibfnamefont {J.}~\bibnamefont {Blazek}}, \bibinfo {author} {\bibfnamefont {B.}~\bibnamefont {Benson}}, \bibinfo {author} {\bibfnamefont {R.}~\bibnamefont {Bernstein}}, \bibinfo {author} {\bibfnamefont {G.}~\bibnamefont {Bernstein}}, \bibinfo {author} {\bibfnamefont {E.}~\bibnamefont {Bertin}}, \bibinfo {author} {\bibfnamefont {D.}~\bibnamefont {Brooks}},  \emph {et~al.},\ }\href@noop {} {\bibfield  {journal} {\bibinfo  {journal} {Monthly Notices of the Royal Astronomical Society}\ }\textbf {\bibinfo {volume} {480}},\ \bibinfo {pages} {3879} (\bibinfo {year} {2018})}\BibitemShut {NoStop}%
\bibitem [{\citenamefont {Riess}\ \emph {et~al.}(2022)\citenamefont {Riess}, \citenamefont {Yuan}, \citenamefont {Macri}, \citenamefont {Scolnic}, \citenamefont {Brout}, \citenamefont {Casertano}, \citenamefont {Jones}, \citenamefont {Murakami}, \citenamefont {Anand}, \citenamefont {Breuval} \emph {et~al.}}]{SH0ES2022}%
  \BibitemOpen
  \bibfield  {author} {\bibinfo {author} {\bibfnamefont {A.~G.}\ \bibnamefont {Riess}}, \bibinfo {author} {\bibfnamefont {W.}~\bibnamefont {Yuan}}, \bibinfo {author} {\bibfnamefont {L.~M.}\ \bibnamefont {Macri}}, \bibinfo {author} {\bibfnamefont {D.}~\bibnamefont {Scolnic}}, \bibinfo {author} {\bibfnamefont {D.}~\bibnamefont {Brout}}, \bibinfo {author} {\bibfnamefont {S.}~\bibnamefont {Casertano}}, \bibinfo {author} {\bibfnamefont {D.~O.}\ \bibnamefont {Jones}}, \bibinfo {author} {\bibfnamefont {Y.}~\bibnamefont {Murakami}}, \bibinfo {author} {\bibfnamefont {G.~S.}\ \bibnamefont {Anand}}, \bibinfo {author} {\bibfnamefont {L.}~\bibnamefont {Breuval}},  \emph {et~al.},\ }\href@noop {} {\bibfield  {journal} {\bibinfo  {journal} {The Astrophysical journal letters}\ }\textbf {\bibinfo {volume} {934}},\ \bibinfo {pages} {L7} (\bibinfo {year} {2022})}\BibitemShut {NoStop}%
\bibitem [{\citenamefont {Verde}\ \emph {et~al.}(2019)\citenamefont {Verde}, \citenamefont {Treu},\ and\ \citenamefont {Riess}}]{verde2019tensionof5sig}%
  \BibitemOpen
  \bibfield  {author} {\bibinfo {author} {\bibfnamefont {L.}~\bibnamefont {Verde}}, \bibinfo {author} {\bibfnamefont {T.}~\bibnamefont {Treu}}, \ and\ \bibinfo {author} {\bibfnamefont {A.~G.}\ \bibnamefont {Riess}},\ }\href@noop {} {\bibfield  {journal} {\bibinfo  {journal} {Nature Astronomy}\ }\textbf {\bibinfo {volume} {3}},\ \bibinfo {pages} {891} (\bibinfo {year} {2019})}\BibitemShut {NoStop}%
\bibitem [{\citenamefont {Abdalla}\ \emph {et~al.}(2022{\natexlab{a}})\citenamefont {Abdalla}, \citenamefont {Abell{\'a}n}, \citenamefont {Aboubrahim}, \citenamefont {Agnello}, \citenamefont {Akarsu}, \citenamefont {Akrami}, \citenamefont {Alestas}, \citenamefont {Aloni}, \citenamefont {Amendola}, \citenamefont {Anchordoqui} \emph {et~al.}}]{abdalla2022cosmologyreview}%
  \BibitemOpen
  \bibfield  {author} {\bibinfo {author} {\bibfnamefont {E.}~\bibnamefont {Abdalla}}, \bibinfo {author} {\bibfnamefont {G.~F.}\ \bibnamefont {Abell{\'a}n}}, \bibinfo {author} {\bibfnamefont {A.}~\bibnamefont {Aboubrahim}}, \bibinfo {author} {\bibfnamefont {A.}~\bibnamefont {Agnello}}, \bibinfo {author} {\bibfnamefont {{\"O}.}~\bibnamefont {Akarsu}}, \bibinfo {author} {\bibfnamefont {Y.}~\bibnamefont {Akrami}}, \bibinfo {author} {\bibfnamefont {G.}~\bibnamefont {Alestas}}, \bibinfo {author} {\bibfnamefont {D.}~\bibnamefont {Aloni}}, \bibinfo {author} {\bibfnamefont {L.}~\bibnamefont {Amendola}}, \bibinfo {author} {\bibfnamefont {L.~A.}\ \bibnamefont {Anchordoqui}},  \emph {et~al.},\ }\href@noop {} {\bibfield  {journal} {\bibinfo  {journal} {Journal of High Energy Astrophysics}\ }\textbf {\bibinfo {volume} {34}},\ \bibinfo {pages} {49} (\bibinfo {year} {2022}{\natexlab{a}})}\BibitemShut {NoStop}%
\bibitem [{\citenamefont {Cuceu}\ \emph {et~al.}(2019)\citenamefont {Cuceu}, \citenamefont {Farr}, \citenamefont {Lemos},\ and\ \citenamefont {Font-Ribera}}]{cuceu2019bao+bbn1}%
  \BibitemOpen
  \bibfield  {author} {\bibinfo {author} {\bibfnamefont {A.}~\bibnamefont {Cuceu}}, \bibinfo {author} {\bibfnamefont {J.}~\bibnamefont {Farr}}, \bibinfo {author} {\bibfnamefont {P.}~\bibnamefont {Lemos}}, \ and\ \bibinfo {author} {\bibfnamefont {A.}~\bibnamefont {Font-Ribera}},\ }\href@noop {} {\bibfield  {journal} {\bibinfo  {journal} {Journal of Cosmology and Astroparticle Physics}\ }\textbf {\bibinfo {volume} {2019}},\ \bibinfo {pages} {044} (\bibinfo {year} {2019})}\BibitemShut {NoStop}%
\bibitem [{\citenamefont {Sch{\"o}neberg}\ \emph {et~al.}(2022)\citenamefont {Sch{\"o}neberg}, \citenamefont {Verde}, \citenamefont {Gil-Mar{\'\i}n},\ and\ \citenamefont {Brieden}}]{schoneberg2022bao+bbn2}%
  \BibitemOpen
  \bibfield  {author} {\bibinfo {author} {\bibfnamefont {N.}~\bibnamefont {Sch{\"o}neberg}}, \bibinfo {author} {\bibfnamefont {L.}~\bibnamefont {Verde}}, \bibinfo {author} {\bibfnamefont {H.}~\bibnamefont {Gil-Mar{\'\i}n}}, \ and\ \bibinfo {author} {\bibfnamefont {S.}~\bibnamefont {Brieden}},\ }\href@noop {} {\bibfield  {journal} {\bibinfo  {journal} {Journal of Cosmology and Astroparticle Physics}\ }\textbf {\bibinfo {volume} {2022}},\ \bibinfo {pages} {039} (\bibinfo {year} {2022})}\BibitemShut {NoStop}%
\bibitem [{\citenamefont {Cooke}\ \emph {et~al.}(2018)\citenamefont {Cooke}, \citenamefont {Pettini},\ and\ \citenamefont {Steidel}}]{cooke2018bbn1}%
  \BibitemOpen
  \bibfield  {author} {\bibinfo {author} {\bibfnamefont {R.~J.}\ \bibnamefont {Cooke}}, \bibinfo {author} {\bibfnamefont {M.}~\bibnamefont {Pettini}}, \ and\ \bibinfo {author} {\bibfnamefont {C.~C.}\ \bibnamefont {Steidel}},\ }\href@noop {} {\bibfield  {journal} {\bibinfo  {journal} {The Astrophysical Journal}\ }\textbf {\bibinfo {volume} {855}},\ \bibinfo {pages} {102} (\bibinfo {year} {2018})}\BibitemShut {NoStop}%
\bibitem [{\citenamefont {Mossa}\ \emph {et~al.}(2020)\citenamefont {Mossa}, \citenamefont {St{\"o}ckel}, \citenamefont {Cavanna}, \citenamefont {Ferraro}, \citenamefont {Aliotta}, \citenamefont {Barile}, \citenamefont {Bemmerer}, \citenamefont {Best}, \citenamefont {Boeltzig}, \citenamefont {Broggini} \emph {et~al.}}]{mossa2020bbn2}%
  \BibitemOpen
  \bibfield  {author} {\bibinfo {author} {\bibfnamefont {V.}~\bibnamefont {Mossa}}, \bibinfo {author} {\bibfnamefont {K.}~\bibnamefont {St{\"o}ckel}}, \bibinfo {author} {\bibfnamefont {F.}~\bibnamefont {Cavanna}}, \bibinfo {author} {\bibfnamefont {F.}~\bibnamefont {Ferraro}}, \bibinfo {author} {\bibfnamefont {M.}~\bibnamefont {Aliotta}}, \bibinfo {author} {\bibfnamefont {F.}~\bibnamefont {Barile}}, \bibinfo {author} {\bibfnamefont {D.}~\bibnamefont {Bemmerer}}, \bibinfo {author} {\bibfnamefont {A.}~\bibnamefont {Best}}, \bibinfo {author} {\bibfnamefont {A.}~\bibnamefont {Boeltzig}}, \bibinfo {author} {\bibfnamefont {C.}~\bibnamefont {Broggini}},  \emph {et~al.},\ }\href@noop {} {\bibfield  {journal} {\bibinfo  {journal} {Nature}\ }\textbf {\bibinfo {volume} {587}},\ \bibinfo {pages} {210} (\bibinfo {year} {2020})}\BibitemShut {NoStop}%
\bibitem [{\citenamefont {Jang}\ and\ \citenamefont {Lee}(2017)}]{jang2017trgb1}%
  \BibitemOpen
  \bibfield  {author} {\bibinfo {author} {\bibfnamefont {I.~S.}\ \bibnamefont {Jang}}\ and\ \bibinfo {author} {\bibfnamefont {M.~G.}\ \bibnamefont {Lee}},\ }\href@noop {} {\bibfield  {journal} {\bibinfo  {journal} {The Astrophysical Journal}\ }\textbf {\bibinfo {volume} {836}},\ \bibinfo {pages} {74} (\bibinfo {year} {2017})}\BibitemShut {NoStop}%
\bibitem [{\citenamefont {Freedman}\ \emph {et~al.}(2020)\citenamefont {Freedman}, \citenamefont {Madore}, \citenamefont {Hoyt}, \citenamefont {Jang}, \citenamefont {Beaton}, \citenamefont {Lee}, \citenamefont {Monson}, \citenamefont {Neeley},\ and\ \citenamefont {Rich}}]{freedman2020trgb2}%
  \BibitemOpen
  \bibfield  {author} {\bibinfo {author} {\bibfnamefont {W.~L.}\ \bibnamefont {Freedman}}, \bibinfo {author} {\bibfnamefont {B.~F.}\ \bibnamefont {Madore}}, \bibinfo {author} {\bibfnamefont {T.}~\bibnamefont {Hoyt}}, \bibinfo {author} {\bibfnamefont {I.~S.}\ \bibnamefont {Jang}}, \bibinfo {author} {\bibfnamefont {R.}~\bibnamefont {Beaton}}, \bibinfo {author} {\bibfnamefont {M.~G.}\ \bibnamefont {Lee}}, \bibinfo {author} {\bibfnamefont {A.}~\bibnamefont {Monson}}, \bibinfo {author} {\bibfnamefont {J.}~\bibnamefont {Neeley}}, \ and\ \bibinfo {author} {\bibfnamefont {J.}~\bibnamefont {Rich}},\ }\href@noop {} {\bibfield  {journal} {\bibinfo  {journal} {The Astrophysical Journal}\ }\textbf {\bibinfo {volume} {891}},\ \bibinfo {pages} {57} (\bibinfo {year} {2020})}\BibitemShut {NoStop}%
\bibitem [{\citenamefont {Huang}\ \emph {et~al.}(2020)\citenamefont {Huang}, \citenamefont {Riess}, \citenamefont {Yuan}, \citenamefont {Macri}, \citenamefont {Zakamska}, \citenamefont {Casertano}, \citenamefont {Whitelock}, \citenamefont {Hoffmann}, \citenamefont {Filippenko},\ and\ \citenamefont {Scolnic}}]{huang2020miras}%
  \BibitemOpen
  \bibfield  {author} {\bibinfo {author} {\bibfnamefont {C.~D.}\ \bibnamefont {Huang}}, \bibinfo {author} {\bibfnamefont {A.~G.}\ \bibnamefont {Riess}}, \bibinfo {author} {\bibfnamefont {W.}~\bibnamefont {Yuan}}, \bibinfo {author} {\bibfnamefont {L.~M.}\ \bibnamefont {Macri}}, \bibinfo {author} {\bibfnamefont {N.~L.}\ \bibnamefont {Zakamska}}, \bibinfo {author} {\bibfnamefont {S.}~\bibnamefont {Casertano}}, \bibinfo {author} {\bibfnamefont {P.~A.}\ \bibnamefont {Whitelock}}, \bibinfo {author} {\bibfnamefont {S.~L.}\ \bibnamefont {Hoffmann}}, \bibinfo {author} {\bibfnamefont {A.~V.}\ \bibnamefont {Filippenko}}, \ and\ \bibinfo {author} {\bibfnamefont {D.}~\bibnamefont {Scolnic}},\ }\href@noop {} {\bibfield  {journal} {\bibinfo  {journal} {The Astrophysical Journal}\ }\textbf {\bibinfo {volume} {889}},\ \bibinfo {pages} {5} (\bibinfo {year} {2020})}\BibitemShut {NoStop}%
\bibitem [{\citenamefont {Kourkchi}\ \emph {et~al.}(2020)\citenamefont {Kourkchi}, \citenamefont {Tully}, \citenamefont {Anand}, \citenamefont {Courtois}, \citenamefont {Dupuy}, \citenamefont {Neill}, \citenamefont {Rizzi},\ and\ \citenamefont {Seibert}}]{kourkchi2020tully-fisher}%
  \BibitemOpen
  \bibfield  {author} {\bibinfo {author} {\bibfnamefont {E.}~\bibnamefont {Kourkchi}}, \bibinfo {author} {\bibfnamefont {R.~B.}\ \bibnamefont {Tully}}, \bibinfo {author} {\bibfnamefont {G.~S.}\ \bibnamefont {Anand}}, \bibinfo {author} {\bibfnamefont {H.~M.}\ \bibnamefont {Courtois}}, \bibinfo {author} {\bibfnamefont {A.}~\bibnamefont {Dupuy}}, \bibinfo {author} {\bibfnamefont {J.~D.}\ \bibnamefont {Neill}}, \bibinfo {author} {\bibfnamefont {L.}~\bibnamefont {Rizzi}}, \ and\ \bibinfo {author} {\bibfnamefont {M.}~\bibnamefont {Seibert}},\ }\href@noop {} {\bibfield  {journal} {\bibinfo  {journal} {The Astrophysical Journal}\ }\textbf {\bibinfo {volume} {896}},\ \bibinfo {pages} {3} (\bibinfo {year} {2020})}\BibitemShut {NoStop}%
\bibitem [{\citenamefont {{Santos}}\ \emph {et~al.}(2017)\citenamefont {{Santos}}, \citenamefont {{Zhao}}, \citenamefont {{Ferreira}},\ and\ \citenamefont {{Quintin}}}]{2017PhRvD..96j3529S}%
  \BibitemOpen
  \bibfield  {author} {\bibinfo {author} {\bibfnamefont {L.}~\bibnamefont {{Santos}}}, \bibinfo {author} {\bibfnamefont {W.}~\bibnamefont {{Zhao}}}, \bibinfo {author} {\bibfnamefont {E.~G.~M.}\ \bibnamefont {{Ferreira}}}, \ and\ \bibinfo {author} {\bibfnamefont {J.}~\bibnamefont {{Quintin}}},\ }\href {\doibase 10.1103/PhysRevD.96.103529} {\bibfield  {journal} {\bibinfo  {journal} {\prd}\ }\textbf {\bibinfo {volume} {96}},\ \bibinfo {eid} {103529} (\bibinfo {year} {2017})},\ \Eprint {http://arxiv.org/abs/1707.06827} {arXiv:1707.06827 [astro-ph.CO]} \BibitemShut {NoStop}%
\bibitem [{\citenamefont {Wong}\ \emph {et~al.}(2020)\citenamefont {Wong}, \citenamefont {Suyu}, \citenamefont {Chen}, \citenamefont {Rusu}, \citenamefont {Millon}, \citenamefont {Sluse}, \citenamefont {Bonvin}, \citenamefont {Fassnacht}, \citenamefont {Taubenberger}, \citenamefont {Auger} \emph {et~al.}}]{wong2020h0licow}%
  \BibitemOpen
  \bibfield  {author} {\bibinfo {author} {\bibfnamefont {K.~C.}\ \bibnamefont {Wong}}, \bibinfo {author} {\bibfnamefont {S.~H.}\ \bibnamefont {Suyu}}, \bibinfo {author} {\bibfnamefont {G.~C.}\ \bibnamefont {Chen}}, \bibinfo {author} {\bibfnamefont {C.~E.}\ \bibnamefont {Rusu}}, \bibinfo {author} {\bibfnamefont {M.}~\bibnamefont {Millon}}, \bibinfo {author} {\bibfnamefont {D.}~\bibnamefont {Sluse}}, \bibinfo {author} {\bibfnamefont {V.}~\bibnamefont {Bonvin}}, \bibinfo {author} {\bibfnamefont {C.~D.}\ \bibnamefont {Fassnacht}}, \bibinfo {author} {\bibfnamefont {S.}~\bibnamefont {Taubenberger}}, \bibinfo {author} {\bibfnamefont {M.~W.}\ \bibnamefont {Auger}},  \emph {et~al.},\ }\href@noop {} {\bibfield  {journal} {\bibinfo  {journal} {Monthly Notices of the Royal Astronomical Society}\ }\textbf {\bibinfo {volume} {498}},\ \bibinfo {pages} {1420} (\bibinfo {year} {2020})}\BibitemShut {NoStop}%
\bibitem [{\citenamefont {Di~Valentino}\ \emph {et~al.}(2021)\citenamefont {Di~Valentino}, \citenamefont {Mena}, \citenamefont {Pan}, \citenamefont {Visinelli}, \citenamefont {Yang}, \citenamefont {Melchiorri}, \citenamefont {Mota}, \citenamefont {Riess},\ and\ \citenamefont {Silk}}]{di2021systemerror}%
  \BibitemOpen
  \bibfield  {author} {\bibinfo {author} {\bibfnamefont {E.}~\bibnamefont {Di~Valentino}}, \bibinfo {author} {\bibfnamefont {O.}~\bibnamefont {Mena}}, \bibinfo {author} {\bibfnamefont {S.}~\bibnamefont {Pan}}, \bibinfo {author} {\bibfnamefont {L.}~\bibnamefont {Visinelli}}, \bibinfo {author} {\bibfnamefont {W.}~\bibnamefont {Yang}}, \bibinfo {author} {\bibfnamefont {A.}~\bibnamefont {Melchiorri}}, \bibinfo {author} {\bibfnamefont {D.~F.}\ \bibnamefont {Mota}}, \bibinfo {author} {\bibfnamefont {A.~G.}\ \bibnamefont {Riess}}, \ and\ \bibinfo {author} {\bibfnamefont {J.}~\bibnamefont {Silk}},\ }\href@noop {} {\bibfield  {journal} {\bibinfo  {journal} {Classical and Quantum Gravity}\ }\textbf {\bibinfo {volume} {38}},\ \bibinfo {pages} {153001} (\bibinfo {year} {2021})}\BibitemShut {NoStop}%
\bibitem [{\citenamefont {Lemos}\ \emph {et~al.}(2019)\citenamefont {Lemos}, \citenamefont {Lee}, \citenamefont {Efstathiou},\ and\ \citenamefont {Gratton}}]{lemos2019inverseladder}%
  \BibitemOpen
  \bibfield  {author} {\bibinfo {author} {\bibfnamefont {P.}~\bibnamefont {Lemos}}, \bibinfo {author} {\bibfnamefont {E.}~\bibnamefont {Lee}}, \bibinfo {author} {\bibfnamefont {G.}~\bibnamefont {Efstathiou}}, \ and\ \bibinfo {author} {\bibfnamefont {S.}~\bibnamefont {Gratton}},\ }\href@noop {} {\bibfield  {journal} {\bibinfo  {journal} {Monthly Notices of the Royal Astronomical Society}\ }\textbf {\bibinfo {volume} {483}},\ \bibinfo {pages} {4803} (\bibinfo {year} {2019})}\BibitemShut {NoStop}%
\bibitem [{\citenamefont {Efstathiou}(2021)}]{efstathiou2021inverseladder2}%
  \BibitemOpen
  \bibfield  {author} {\bibinfo {author} {\bibfnamefont {G.}~\bibnamefont {Efstathiou}},\ }\href@noop {} {\bibfield  {journal} {\bibinfo  {journal} {Monthly Notices of the Royal Astronomical Society}\ }\textbf {\bibinfo {volume} {505}},\ \bibinfo {pages} {3866} (\bibinfo {year} {2021})}\BibitemShut {NoStop}%
\bibitem [{\citenamefont {Karwal}\ and\ \citenamefont {Kamionkowski}(2016)}]{karwal2016ede1}%
  \BibitemOpen
  \bibfield  {author} {\bibinfo {author} {\bibfnamefont {T.}~\bibnamefont {Karwal}}\ and\ \bibinfo {author} {\bibfnamefont {M.}~\bibnamefont {Kamionkowski}},\ }\href@noop {} {\bibfield  {journal} {\bibinfo  {journal} {Physical Review D}\ }\textbf {\bibinfo {volume} {94}},\ \bibinfo {pages} {103523} (\bibinfo {year} {2016})}\BibitemShut {NoStop}%
\bibitem [{\citenamefont {M{\"o}rtsell}\ and\ \citenamefont {Dhawan}(2018)}]{mortsell2018ede2}%
  \BibitemOpen
  \bibfield  {author} {\bibinfo {author} {\bibfnamefont {E.}~\bibnamefont {M{\"o}rtsell}}\ and\ \bibinfo {author} {\bibfnamefont {S.}~\bibnamefont {Dhawan}},\ }\href@noop {} {\bibfield  {journal} {\bibinfo  {journal} {Journal of Cosmology and Astroparticle Physics}\ }\textbf {\bibinfo {volume} {2018}},\ \bibinfo {pages} {025} (\bibinfo {year} {2018})}\BibitemShut {NoStop}%
\bibitem [{\citenamefont {Poulin}\ \emph {et~al.}(2019)\citenamefont {Poulin}, \citenamefont {Smith}, \citenamefont {Karwal},\ and\ \citenamefont {Kamionkowski}}]{poulin2019ede3}%
  \BibitemOpen
  \bibfield  {author} {\bibinfo {author} {\bibfnamefont {V.}~\bibnamefont {Poulin}}, \bibinfo {author} {\bibfnamefont {T.~L.}\ \bibnamefont {Smith}}, \bibinfo {author} {\bibfnamefont {T.}~\bibnamefont {Karwal}}, \ and\ \bibinfo {author} {\bibfnamefont {M.}~\bibnamefont {Kamionkowski}},\ }\href@noop {} {\bibfield  {journal} {\bibinfo  {journal} {Physical review letters}\ }\textbf {\bibinfo {volume} {122}},\ \bibinfo {pages} {221301} (\bibinfo {year} {2019})}\BibitemShut {NoStop}%
\bibitem [{\citenamefont {Krishnan}\ \emph {et~al.}(2021)\citenamefont {Krishnan}, \citenamefont {Colg{\'a}in}, \citenamefont {Sheikh-Jabbari},\ and\ \citenamefont {Yang}}]{krishnan2021running}%
  \BibitemOpen
  \bibfield  {author} {\bibinfo {author} {\bibfnamefont {C.}~\bibnamefont {Krishnan}}, \bibinfo {author} {\bibfnamefont {E.~{\'O}.}\ \bibnamefont {Colg{\'a}in}}, \bibinfo {author} {\bibfnamefont {M.}~\bibnamefont {Sheikh-Jabbari}}, \ and\ \bibinfo {author} {\bibfnamefont {T.}~\bibnamefont {Yang}},\ }\href@noop {} {\bibfield  {journal} {\bibinfo  {journal} {Physical Review D}\ }\textbf {\bibinfo {volume} {103}},\ \bibinfo {pages} {103509} (\bibinfo {year} {2021})}\BibitemShut {NoStop}%
\bibitem [{\citenamefont {Colg{\'a}in}\ \emph {et~al.}(2022)\citenamefont {Colg{\'a}in}, \citenamefont {Sheikh-Jabbari}, \citenamefont {Solomon}, \citenamefont {Bargiacchi}, \citenamefont {Capozziello}, \citenamefont {Dainotti},\ and\ \citenamefont {Stojkovic}}]{colgain2022revealing}%
  \BibitemOpen
  \bibfield  {author} {\bibinfo {author} {\bibfnamefont {E.~{\'O}.}\ \bibnamefont {Colg{\'a}in}}, \bibinfo {author} {\bibfnamefont {M.}~\bibnamefont {Sheikh-Jabbari}}, \bibinfo {author} {\bibfnamefont {R.}~\bibnamefont {Solomon}}, \bibinfo {author} {\bibfnamefont {G.}~\bibnamefont {Bargiacchi}}, \bibinfo {author} {\bibfnamefont {S.}~\bibnamefont {Capozziello}}, \bibinfo {author} {\bibfnamefont {M.~G.}\ \bibnamefont {Dainotti}}, \ and\ \bibinfo {author} {\bibfnamefont {D.}~\bibnamefont {Stojkovic}},\ }\href@noop {} {\bibfield  {journal} {\bibinfo  {journal} {Physical Review D}\ }\textbf {\bibinfo {volume} {106}},\ \bibinfo {pages} {L041301} (\bibinfo {year} {2022})}\BibitemShut {NoStop}%
\bibitem [{\citenamefont {Malekjani}\ \emph {et~al.}(2023)\citenamefont {Malekjani}, \citenamefont {Conville}, \citenamefont {Colg{\'a}in}, \citenamefont {Pourojaghi},\ and\ \citenamefont {Sheikh-Jabbari}}]{malekjani2023negative}%
  \BibitemOpen
  \bibfield  {author} {\bibinfo {author} {\bibfnamefont {M.}~\bibnamefont {Malekjani}}, \bibinfo {author} {\bibfnamefont {R.~M.}\ \bibnamefont {Conville}}, \bibinfo {author} {\bibfnamefont {E.~{\'O}.}\ \bibnamefont {Colg{\'a}in}}, \bibinfo {author} {\bibfnamefont {S.}~\bibnamefont {Pourojaghi}}, \ and\ \bibinfo {author} {\bibfnamefont {M.}~\bibnamefont {Sheikh-Jabbari}},\ }\href@noop {} {\bibfield  {journal} {\bibinfo  {journal} {arXiv preprint arXiv:2301.12725}\ } (\bibinfo {year} {2023})}\BibitemShut {NoStop}%
\bibitem [{\citenamefont {Eisenstein}\ and\ \citenamefont {Hu}(1998)}]{eisenstein1998baorigin}%
  \BibitemOpen
  \bibfield  {author} {\bibinfo {author} {\bibfnamefont {D.~J.}\ \bibnamefont {Eisenstein}}\ and\ \bibinfo {author} {\bibfnamefont {W.}~\bibnamefont {Hu}},\ }\href@noop {} {\bibfield  {journal} {\bibinfo  {journal} {The Astrophysical Journal}\ }\textbf {\bibinfo {volume} {496}},\ \bibinfo {pages} {605} (\bibinfo {year} {1998})}\BibitemShut {NoStop}%
\bibitem [{\citenamefont {Hogg}(1999)}]{hogg1999distance}%
  \BibitemOpen
  \bibfield  {author} {\bibinfo {author} {\bibfnamefont {D.~W.}\ \bibnamefont {Hogg}},\ }\href@noop {} {\bibfield  {journal} {\bibinfo  {journal} {arXiv preprint astro-ph/9905116}\ } (\bibinfo {year} {1999})}\BibitemShut {NoStop}%
\bibitem [{\citenamefont {Hu}\ and\ \citenamefont {Sugiyama}(1996)}]{hu1996analyticalformula}%
  \BibitemOpen
  \bibfield  {author} {\bibinfo {author} {\bibfnamefont {W.}~\bibnamefont {Hu}}\ and\ \bibinfo {author} {\bibfnamefont {N.}~\bibnamefont {Sugiyama}},\ }\href@noop {} {\enquote {\bibinfo {title} {Small scale cosmological perturbations: An analytic approach https://doi. org/10.1086/177989 astrophys},}\ } (\bibinfo {year} {1996})\BibitemShut {NoStop}%
\bibitem [{\citenamefont {Komatsu}\ \emph {et~al.}(2009)\citenamefont {Komatsu}, \citenamefont {Dunkley}, \citenamefont {Nolta}, \citenamefont {Bennett}, \citenamefont {Gold}, \citenamefont {Hinshaw}, \citenamefont {Jarosik}, \citenamefont {Larson}, \citenamefont {Limon}, \citenamefont {Page} \emph {et~al.}}]{komatsu2009five}%
  \BibitemOpen
  \bibfield  {author} {\bibinfo {author} {\bibfnamefont {E.}~\bibnamefont {Komatsu}}, \bibinfo {author} {\bibfnamefont {J.}~\bibnamefont {Dunkley}}, \bibinfo {author} {\bibfnamefont {M.}~\bibnamefont {Nolta}}, \bibinfo {author} {\bibfnamefont {C.}~\bibnamefont {Bennett}}, \bibinfo {author} {\bibfnamefont {B.}~\bibnamefont {Gold}}, \bibinfo {author} {\bibfnamefont {G.}~\bibnamefont {Hinshaw}}, \bibinfo {author} {\bibfnamefont {N.}~\bibnamefont {Jarosik}}, \bibinfo {author} {\bibfnamefont {D.}~\bibnamefont {Larson}}, \bibinfo {author} {\bibfnamefont {M.}~\bibnamefont {Limon}}, \bibinfo {author} {\bibfnamefont {L.}~\bibnamefont {Page}},  \emph {et~al.},\ }\href@noop {} {\bibfield  {journal} {\bibinfo  {journal} {The Astrophysical Journal Supplement Series}\ }\textbf {\bibinfo {volume} {180}},\ \bibinfo {pages} {330} (\bibinfo {year} {2009})}\BibitemShut {NoStop}%
\bibitem [{\citenamefont {Wang}\ and\ \citenamefont {Mukherjee}(2007)}]{wang2007distanceprior-1}%
  \BibitemOpen
  \bibfield  {author} {\bibinfo {author} {\bibfnamefont {Y.}~\bibnamefont {Wang}}\ and\ \bibinfo {author} {\bibfnamefont {P.}~\bibnamefont {Mukherjee}},\ }\href@noop {} {\bibfield  {journal} {\bibinfo  {journal} {Physical Review D}\ }\textbf {\bibinfo {volume} {76}},\ \bibinfo {pages} {103533} (\bibinfo {year} {2007})}\BibitemShut {NoStop}%
\bibitem [{\citenamefont {Wang}\ and\ \citenamefont {Wang}(2013)}]{wang2013distanceprior-2}%
  \BibitemOpen
  \bibfield  {author} {\bibinfo {author} {\bibfnamefont {Y.}~\bibnamefont {Wang}}\ and\ \bibinfo {author} {\bibfnamefont {S.}~\bibnamefont {Wang}},\ }\href@noop {} {\bibfield  {journal} {\bibinfo  {journal} {Physical Review D}\ }\textbf {\bibinfo {volume} {88}},\ \bibinfo {pages} {043522} (\bibinfo {year} {2013})}\BibitemShut {NoStop}%
\bibitem [{\citenamefont {Zhai}\ and\ \citenamefont {Wang}(2019)}]{wang2019distanceprior-3}%
  \BibitemOpen
  \bibfield  {author} {\bibinfo {author} {\bibfnamefont {Z.}~\bibnamefont {Zhai}}\ and\ \bibinfo {author} {\bibfnamefont {Y.}~\bibnamefont {Wang}},\ }\href@noop {} {\bibfield  {journal} {\bibinfo  {journal} {Journal of Cosmology and Astroparticle Physics}\ }\textbf {\bibinfo {volume} {2019}},\ \bibinfo {pages} {005} (\bibinfo {year} {2019})}\BibitemShut {NoStop}%
\bibitem [{\citenamefont {Beutler}\ \emph {et~al.}(2011)\citenamefont {Beutler}, \citenamefont {Blake}, \citenamefont {Colless}, \citenamefont {Jones}, \citenamefont {Staveley-Smith}, \citenamefont {Campbell}, \citenamefont {Parker}, \citenamefont {Saunders},\ and\ \citenamefont {Watson}}]{beutler20116df}%
  \BibitemOpen
  \bibfield  {author} {\bibinfo {author} {\bibfnamefont {F.}~\bibnamefont {Beutler}}, \bibinfo {author} {\bibfnamefont {C.}~\bibnamefont {Blake}}, \bibinfo {author} {\bibfnamefont {M.}~\bibnamefont {Colless}}, \bibinfo {author} {\bibfnamefont {D.~H.}\ \bibnamefont {Jones}}, \bibinfo {author} {\bibfnamefont {L.}~\bibnamefont {Staveley-Smith}}, \bibinfo {author} {\bibfnamefont {L.}~\bibnamefont {Campbell}}, \bibinfo {author} {\bibfnamefont {Q.}~\bibnamefont {Parker}}, \bibinfo {author} {\bibfnamefont {W.}~\bibnamefont {Saunders}}, \ and\ \bibinfo {author} {\bibfnamefont {F.}~\bibnamefont {Watson}},\ }\href@noop {} {\bibfield  {journal} {\bibinfo  {journal} {Monthly Notices of the Royal Astronomical Society}\ }\textbf {\bibinfo {volume} {416}},\ \bibinfo {pages} {3017} (\bibinfo {year} {2011})}\BibitemShut {NoStop}%
\bibitem [{\citenamefont {Ross}\ \emph {et~al.}(2015)\citenamefont {Ross}, \citenamefont {Samushia}, \citenamefont {Howlett}, \citenamefont {Percival}, \citenamefont {Burden},\ and\ \citenamefont {Manera}}]{ross2015mgs}%
  \BibitemOpen
  \bibfield  {author} {\bibinfo {author} {\bibfnamefont {A.~J.}\ \bibnamefont {Ross}}, \bibinfo {author} {\bibfnamefont {L.}~\bibnamefont {Samushia}}, \bibinfo {author} {\bibfnamefont {C.}~\bibnamefont {Howlett}}, \bibinfo {author} {\bibfnamefont {W.~J.}\ \bibnamefont {Percival}}, \bibinfo {author} {\bibfnamefont {A.}~\bibnamefont {Burden}}, \ and\ \bibinfo {author} {\bibfnamefont {M.}~\bibnamefont {Manera}},\ }\href@noop {} {\bibfield  {journal} {\bibinfo  {journal} {Monthly Notices of the Royal Astronomical Society}\ }\textbf {\bibinfo {volume} {449}},\ \bibinfo {pages} {835} (\bibinfo {year} {2015})}\BibitemShut {NoStop}%
\bibitem [{\citenamefont {Des~Bourboux}\ \emph {et~al.}(2020)\citenamefont {Des~Bourboux}, \citenamefont {Rich}, \citenamefont {Font-Ribera}, \citenamefont {de~Sainte~Agathe}, \citenamefont {Farr}, \citenamefont {Etourneau}, \citenamefont {Le~Goff}, \citenamefont {Cuceu}, \citenamefont {Balland}, \citenamefont {Bautista} \emph {et~al.}}]{bouroux2020lya}%
  \BibitemOpen
  \bibfield  {author} {\bibinfo {author} {\bibfnamefont {H.~D.~M.}\ \bibnamefont {Des~Bourboux}}, \bibinfo {author} {\bibfnamefont {J.}~\bibnamefont {Rich}}, \bibinfo {author} {\bibfnamefont {A.}~\bibnamefont {Font-Ribera}}, \bibinfo {author} {\bibfnamefont {V.}~\bibnamefont {de~Sainte~Agathe}}, \bibinfo {author} {\bibfnamefont {J.}~\bibnamefont {Farr}}, \bibinfo {author} {\bibfnamefont {T.}~\bibnamefont {Etourneau}}, \bibinfo {author} {\bibfnamefont {J.-M.}\ \bibnamefont {Le~Goff}}, \bibinfo {author} {\bibfnamefont {A.}~\bibnamefont {Cuceu}}, \bibinfo {author} {\bibfnamefont {C.}~\bibnamefont {Balland}}, \bibinfo {author} {\bibfnamefont {J.~E.}\ \bibnamefont {Bautista}},  \emph {et~al.},\ }\href@noop {} {\bibfield  {journal} {\bibinfo  {journal} {The Astrophysical Journal}\ }\textbf {\bibinfo {volume} {901}},\ \bibinfo {pages} {153} (\bibinfo {year} {2020})}\BibitemShut {NoStop}%
\bibitem [{\citenamefont {De~Mattia}\ \emph {et~al.}(2021)\citenamefont {De~Mattia}, \citenamefont {Ruhlmann-Kleider}, \citenamefont {Raichoor}, \citenamefont {Ross}, \citenamefont {Tamone}, \citenamefont {Zhao}, \citenamefont {Alam}, \citenamefont {Avila}, \citenamefont {Burtin}, \citenamefont {Bautista} \emph {et~al.}}]{de2021elg}%
  \BibitemOpen
  \bibfield  {author} {\bibinfo {author} {\bibfnamefont {A.}~\bibnamefont {De~Mattia}}, \bibinfo {author} {\bibfnamefont {V.}~\bibnamefont {Ruhlmann-Kleider}}, \bibinfo {author} {\bibfnamefont {A.}~\bibnamefont {Raichoor}}, \bibinfo {author} {\bibfnamefont {A.~J.}\ \bibnamefont {Ross}}, \bibinfo {author} {\bibfnamefont {A.}~\bibnamefont {Tamone}}, \bibinfo {author} {\bibfnamefont {C.}~\bibnamefont {Zhao}}, \bibinfo {author} {\bibfnamefont {S.}~\bibnamefont {Alam}}, \bibinfo {author} {\bibfnamefont {S.}~\bibnamefont {Avila}}, \bibinfo {author} {\bibfnamefont {E.}~\bibnamefont {Burtin}}, \bibinfo {author} {\bibfnamefont {J.}~\bibnamefont {Bautista}},  \emph {et~al.},\ }\href@noop {} {\bibfield  {journal} {\bibinfo  {journal} {Monthly Notices of the Royal Astronomical Society}\ }\textbf {\bibinfo {volume} {501}},\ \bibinfo {pages} {5616} (\bibinfo {year} {2021})}\BibitemShut {NoStop}%
\bibitem [{\citenamefont {{Foreman-Mackey}}\ \emph {et~al.}(2013)\citenamefont {{Foreman-Mackey}}, \citenamefont {{Hogg}}, \citenamefont {{Lang}},\ and\ \citenamefont {{Goodman}}}]{emcee}%
  \BibitemOpen
  \bibfield  {author} {\bibinfo {author} {\bibfnamefont {D.}~\bibnamefont {{Foreman-Mackey}}}, \bibinfo {author} {\bibfnamefont {D.~W.}\ \bibnamefont {{Hogg}}}, \bibinfo {author} {\bibfnamefont {D.}~\bibnamefont {{Lang}}}, \ and\ \bibinfo {author} {\bibfnamefont {J.}~\bibnamefont {{Goodman}}},\ }\href {\doibase 10.1086/670067} {\bibfield  {journal} {\bibinfo  {journal} {PASP}\ }\textbf {\bibinfo {volume} {125}},\ \bibinfo {pages} {306} (\bibinfo {year} {2013})},\ \Eprint {http://arxiv.org/abs/1202.3665} {1202.3665} \BibitemShut {NoStop}%
\bibitem [{\citenamefont {Lemos}\ and\ \citenamefont {Lewis}(2023)}]{lemos2023cmbearly}%
  \BibitemOpen
  \bibfield  {author} {\bibinfo {author} {\bibfnamefont {P.}~\bibnamefont {Lemos}}\ and\ \bibinfo {author} {\bibfnamefont {A.}~\bibnamefont {Lewis}},\ }\href@noop {} {\bibfield  {journal} {\bibinfo  {journal} {Physical Review D}\ }\textbf {\bibinfo {volume} {107}},\ \bibinfo {pages} {103505} (\bibinfo {year} {2023})}\BibitemShut {NoStop}%
\bibitem [{\citenamefont {Pogosian}\ \emph {et~al.}(2020)\citenamefont {Pogosian}, \citenamefont {Zhao},\ and\ \citenamefont {Jedamzik}}]{pogosian2020baowithprior}%
  \BibitemOpen
  \bibfield  {author} {\bibinfo {author} {\bibfnamefont {L.}~\bibnamefont {Pogosian}}, \bibinfo {author} {\bibfnamefont {G.-B.}\ \bibnamefont {Zhao}}, \ and\ \bibinfo {author} {\bibfnamefont {K.}~\bibnamefont {Jedamzik}},\ }\href@noop {} {\bibfield  {journal} {\bibinfo  {journal} {The Astrophysical Journal Letters}\ }\textbf {\bibinfo {volume} {904}},\ \bibinfo {pages} {L17} (\bibinfo {year} {2020})}\BibitemShut {NoStop}%
\bibitem [{\citenamefont {Freedman}(2021)}]{freedman2021trgb}%
  \BibitemOpen
  \bibfield  {author} {\bibinfo {author} {\bibfnamefont {W.~L.}\ \bibnamefont {Freedman}},\ }\href@noop {} {\bibfield  {journal} {\bibinfo  {journal} {The Astrophysical Journal}\ }\textbf {\bibinfo {volume} {919}},\ \bibinfo {pages} {16} (\bibinfo {year} {2021})}\BibitemShut {NoStop}%
\bibitem [{\citenamefont {Chang}\ \emph {et~al.}(2008)\citenamefont {Chang}, \citenamefont {Pen}, \citenamefont {Peterson},\ and\ \citenamefont {McDonald}}]{chang2008beginofim}%
  \BibitemOpen
  \bibfield  {author} {\bibinfo {author} {\bibfnamefont {T.-C.}\ \bibnamefont {Chang}}, \bibinfo {author} {\bibfnamefont {U.-L.}\ \bibnamefont {Pen}}, \bibinfo {author} {\bibfnamefont {J.~B.}\ \bibnamefont {Peterson}}, \ and\ \bibinfo {author} {\bibfnamefont {P.}~\bibnamefont {McDonald}},\ }\href@noop {} {\bibfield  {journal} {\bibinfo  {journal} {Physical Review Letters}\ }\textbf {\bibinfo {volume} {100}},\ \bibinfo {pages} {091303} (\bibinfo {year} {2008})}\BibitemShut {NoStop}%
\bibitem [{\citenamefont {Blake}\ \emph {et~al.}(2006)\citenamefont {Blake}, \citenamefont {Parkinson}, \citenamefont {Bassett}, \citenamefont {Glazebrook}, \citenamefont {Kunz},\ and\ \citenamefont {Nichol}}]{blake2006formulae}%
  \BibitemOpen
  \bibfield  {author} {\bibinfo {author} {\bibfnamefont {C.}~\bibnamefont {Blake}}, \bibinfo {author} {\bibfnamefont {D.}~\bibnamefont {Parkinson}}, \bibinfo {author} {\bibfnamefont {B.}~\bibnamefont {Bassett}}, \bibinfo {author} {\bibfnamefont {K.}~\bibnamefont {Glazebrook}}, \bibinfo {author} {\bibfnamefont {M.}~\bibnamefont {Kunz}}, \ and\ \bibinfo {author} {\bibfnamefont {R.~C.}\ \bibnamefont {Nichol}},\ }\href@noop {} {\bibfield  {journal} {\bibinfo  {journal} {Monthly Notices of the Royal Astronomical Society}\ }\textbf {\bibinfo {volume} {365}},\ \bibinfo {pages} {255} (\bibinfo {year} {2006})}\BibitemShut {NoStop}%
\bibitem [{\citenamefont {Blake}\ and\ \citenamefont {Glazebrook}(2003)}]{blake2003bg03}%
  \BibitemOpen
  \bibfield  {author} {\bibinfo {author} {\bibfnamefont {C.}~\bibnamefont {Blake}}\ and\ \bibinfo {author} {\bibfnamefont {K.}~\bibnamefont {Glazebrook}},\ }\href@noop {} {\bibfield  {journal} {\bibinfo  {journal} {The Astrophysical Journal}\ }\textbf {\bibinfo {volume} {594}},\ \bibinfo {pages} {665} (\bibinfo {year} {2003})}\BibitemShut {NoStop}%
\bibitem [{\citenamefont {Glazebrook}\ and\ \citenamefont {Blake}(2005)}]{glazebrook2005detailsforbg03}%
  \BibitemOpen
  \bibfield  {author} {\bibinfo {author} {\bibfnamefont {K.}~\bibnamefont {Glazebrook}}\ and\ \bibinfo {author} {\bibfnamefont {C.}~\bibnamefont {Blake}},\ }\href@noop {} {\bibfield  {journal} {\bibinfo  {journal} {The Astrophysical Journal}\ }\textbf {\bibinfo {volume} {631}},\ \bibinfo {pages} {1} (\bibinfo {year} {2005})}\BibitemShut {NoStop}%
\bibitem [{\citenamefont {Peacock}\ and\ \citenamefont {Dodds}(1994)}]{peacock1994reconstructing}%
  \BibitemOpen
  \bibfield  {author} {\bibinfo {author} {\bibfnamefont {J.}~\bibnamefont {Peacock}}\ and\ \bibinfo {author} {\bibfnamefont {S.}~\bibnamefont {Dodds}},\ }\href@noop {} {\bibfield  {journal} {\bibinfo  {journal} {Monthly Notices of the Royal Astronomical Society}\ }\textbf {\bibinfo {volume} {267}},\ \bibinfo {pages} {1020} (\bibinfo {year} {1994})}\BibitemShut {NoStop}%
\bibitem [{\citenamefont {Carroll}\ \emph {et~al.}(1992)\citenamefont {Carroll}, \citenamefont {Press},\ and\ \citenamefont {Turner}}]{carroll1992Dz}%
  \BibitemOpen
  \bibfield  {author} {\bibinfo {author} {\bibfnamefont {S.~M.}\ \bibnamefont {Carroll}}, \bibinfo {author} {\bibfnamefont {W.~H.}\ \bibnamefont {Press}}, \ and\ \bibinfo {author} {\bibfnamefont {E.~L.}\ \bibnamefont {Turner}},\ }\href@noop {} {\bibfield  {journal} {\bibinfo  {journal} {Annual review of astronomy and astrophysics}\ }\textbf {\bibinfo {volume} {30}},\ \bibinfo {pages} {499} (\bibinfo {year} {1992})}\BibitemShut {NoStop}%
\bibitem [{\citenamefont {Richard}\ \emph {et~al.}(2019)\citenamefont {Richard}, \citenamefont {Kneib}, \citenamefont {Blake}, \citenamefont {Raichoor}, \citenamefont {Comparat}, \citenamefont {Shanks}, \citenamefont {Sorce}, \citenamefont {Sahl{\'e}n}, \citenamefont {Howlett}, \citenamefont {Tempel} \emph {et~al.}}]{richard20194most}%
  \BibitemOpen
  \bibfield  {author} {\bibinfo {author} {\bibfnamefont {J.}~\bibnamefont {Richard}}, \bibinfo {author} {\bibfnamefont {J.-P.}\ \bibnamefont {Kneib}}, \bibinfo {author} {\bibfnamefont {C.}~\bibnamefont {Blake}}, \bibinfo {author} {\bibfnamefont {A.}~\bibnamefont {Raichoor}}, \bibinfo {author} {\bibfnamefont {J.}~\bibnamefont {Comparat}}, \bibinfo {author} {\bibfnamefont {T.}~\bibnamefont {Shanks}}, \bibinfo {author} {\bibfnamefont {J.}~\bibnamefont {Sorce}}, \bibinfo {author} {\bibfnamefont {M.}~\bibnamefont {Sahl{\'e}n}}, \bibinfo {author} {\bibfnamefont {C.}~\bibnamefont {Howlett}}, \bibinfo {author} {\bibfnamefont {E.}~\bibnamefont {Tempel}},  \emph {et~al.},\ }\href@noop {} {\bibfield  {journal} {\bibinfo  {journal} {arXiv preprint arXiv:1903.02474}\ } (\bibinfo {year} {2019})}\BibitemShut {NoStop}%
\bibitem [{\citenamefont {Spergel}\ \emph {et~al.}(2013)\citenamefont {Spergel}, \citenamefont {Gehrels}, \citenamefont {Breckinridge}, \citenamefont {Donahue}, \citenamefont {Dressler}, \citenamefont {Gaudi}, \citenamefont {Greene}, \citenamefont {Guyon}, \citenamefont {Hirata}, \citenamefont {Kalirai} \emph {et~al.}}]{spergel2013wfirst}%
  \BibitemOpen
  \bibfield  {author} {\bibinfo {author} {\bibfnamefont {D.}~\bibnamefont {Spergel}}, \bibinfo {author} {\bibfnamefont {N.}~\bibnamefont {Gehrels}}, \bibinfo {author} {\bibfnamefont {J.}~\bibnamefont {Breckinridge}}, \bibinfo {author} {\bibfnamefont {M.}~\bibnamefont {Donahue}}, \bibinfo {author} {\bibfnamefont {A.}~\bibnamefont {Dressler}}, \bibinfo {author} {\bibfnamefont {B.}~\bibnamefont {Gaudi}}, \bibinfo {author} {\bibfnamefont {T.}~\bibnamefont {Greene}}, \bibinfo {author} {\bibfnamefont {O.}~\bibnamefont {Guyon}}, \bibinfo {author} {\bibfnamefont {C.}~\bibnamefont {Hirata}}, \bibinfo {author} {\bibfnamefont {J.}~\bibnamefont {Kalirai}},  \emph {et~al.},\ }\href@noop {} {\bibfield  {journal} {\bibinfo  {journal} {arXiv preprint arXiv:1305.5422}\ } (\bibinfo {year} {2013})}\BibitemShut {NoStop}%
\bibitem [{\citenamefont {Zhan}(2011)}]{zhan2011csst1}%
  \BibitemOpen
  \bibfield  {author} {\bibinfo {author} {\bibfnamefont {H.}~\bibnamefont {Zhan}},\ }\href@noop {} {\bibfield  {journal} {\bibinfo  {journal} {Scientia Sinica Physica, Mechanica \& Astronomica}\ }\textbf {\bibinfo {volume} {41}},\ \bibinfo {pages} {1441} (\bibinfo {year} {2011})}\BibitemShut {NoStop}%
\bibitem [{\citenamefont {Zhan}(2021)}]{zhan2021csst2}%
  \BibitemOpen
  \bibfield  {author} {\bibinfo {author} {\bibfnamefont {H.}~\bibnamefont {Zhan}},\ }\href@noop {} {\bibfield  {journal} {\bibinfo  {journal} {Chinese Science Bulletin}\ }\textbf {\bibinfo {volume} {66}},\ \bibinfo {pages} {1290} (\bibinfo {year} {2021})}\BibitemShut {NoStop}%
\bibitem [{\citenamefont {Laureijs}\ \emph {et~al.}(2011)\citenamefont {Laureijs}, \citenamefont {Amiaux}, \citenamefont {Arduini}, \citenamefont {Augueres}, \citenamefont {Brinchmann}, \citenamefont {Cole}, \citenamefont {Cropper}, \citenamefont {Dabin}, \citenamefont {Duvet}, \citenamefont {Ealet} \emph {et~al.}}]{laureijs2011euclid1}%
  \BibitemOpen
  \bibfield  {author} {\bibinfo {author} {\bibfnamefont {R.}~\bibnamefont {Laureijs}}, \bibinfo {author} {\bibfnamefont {J.}~\bibnamefont {Amiaux}}, \bibinfo {author} {\bibfnamefont {S.}~\bibnamefont {Arduini}}, \bibinfo {author} {\bibfnamefont {J.-L.}\ \bibnamefont {Augueres}}, \bibinfo {author} {\bibfnamefont {J.}~\bibnamefont {Brinchmann}}, \bibinfo {author} {\bibfnamefont {R.}~\bibnamefont {Cole}}, \bibinfo {author} {\bibfnamefont {M.}~\bibnamefont {Cropper}}, \bibinfo {author} {\bibfnamefont {C.}~\bibnamefont {Dabin}}, \bibinfo {author} {\bibfnamefont {L.}~\bibnamefont {Duvet}}, \bibinfo {author} {\bibfnamefont {A.}~\bibnamefont {Ealet}},  \emph {et~al.},\ }\href@noop {} {\bibfield  {journal} {\bibinfo  {journal} {arXiv preprint arXiv:1110.3193}\ } (\bibinfo {year} {2011})}\BibitemShut {NoStop}%
\bibitem [{\citenamefont {Blanchard}\ \emph {et~al.}(2020)\citenamefont {Blanchard}, \citenamefont {Camera}, \citenamefont {Carbone}, \citenamefont {Cardone}, \citenamefont {Casas}, \citenamefont {Clesse}, \citenamefont {Ili{\'c}}, \citenamefont {Kilbinger}, \citenamefont {Kitching}, \citenamefont {Kunz} \emph {et~al.}}]{blanchard2020euclid2}%
  \BibitemOpen
  \bibfield  {author} {\bibinfo {author} {\bibfnamefont {A.}~\bibnamefont {Blanchard}}, \bibinfo {author} {\bibfnamefont {S.}~\bibnamefont {Camera}}, \bibinfo {author} {\bibfnamefont {C.}~\bibnamefont {Carbone}}, \bibinfo {author} {\bibfnamefont {V.}~\bibnamefont {Cardone}}, \bibinfo {author} {\bibfnamefont {S.}~\bibnamefont {Casas}}, \bibinfo {author} {\bibfnamefont {S.}~\bibnamefont {Clesse}}, \bibinfo {author} {\bibfnamefont {S.}~\bibnamefont {Ili{\'c}}}, \bibinfo {author} {\bibfnamefont {M.}~\bibnamefont {Kilbinger}}, \bibinfo {author} {\bibfnamefont {T.}~\bibnamefont {Kitching}}, \bibinfo {author} {\bibfnamefont {M.}~\bibnamefont {Kunz}},  \emph {et~al.},\ }\href@noop {} {\bibfield  {journal} {\bibinfo  {journal} {Astronomy \& Astrophysics}\ }\textbf {\bibinfo {volume} {642}},\ \bibinfo {pages} {A191} (\bibinfo {year} {2020})}\BibitemShut {NoStop}%
\bibitem [{\citenamefont {Font-Ribera}\ \emph {et~al.}(2014)\citenamefont {Font-Ribera}, \citenamefont {McDonald}, \citenamefont {Mostek}, \citenamefont {Reid}, \citenamefont {Seo},\ and\ \citenamefont {Slosar}}]{font2014desi}%
  \BibitemOpen
  \bibfield  {author} {\bibinfo {author} {\bibfnamefont {A.}~\bibnamefont {Font-Ribera}}, \bibinfo {author} {\bibfnamefont {P.}~\bibnamefont {McDonald}}, \bibinfo {author} {\bibfnamefont {N.}~\bibnamefont {Mostek}}, \bibinfo {author} {\bibfnamefont {B.~A.}\ \bibnamefont {Reid}}, \bibinfo {author} {\bibfnamefont {H.-J.}\ \bibnamefont {Seo}}, \ and\ \bibinfo {author} {\bibfnamefont {A.}~\bibnamefont {Slosar}},\ }\href@noop {} {\bibfield  {journal} {\bibinfo  {journal} {Journal of Cosmology and Astroparticle Physics}\ }\textbf {\bibinfo {volume} {2014}},\ \bibinfo {pages} {023} (\bibinfo {year} {2014})}\BibitemShut {NoStop}%
\bibitem [{\citenamefont {Ding}\ \emph {et~al.}(2023)\citenamefont {Ding}, \citenamefont {Yu},\ and\ \citenamefont {Zhang}}]{ding2023csstdata}%
  \BibitemOpen
  \bibfield  {author} {\bibinfo {author} {\bibfnamefont {Z.}~\bibnamefont {Ding}}, \bibinfo {author} {\bibfnamefont {Y.}~\bibnamefont {Yu}}, \ and\ \bibinfo {author} {\bibfnamefont {P.}~\bibnamefont {Zhang}},\ }\href@noop {} {\bibfield  {journal} {\bibinfo  {journal} {arXiv preprint arXiv:2305.00404}\ } (\bibinfo {year} {2023})}\BibitemShut {NoStop}%
\bibitem [{\citenamefont {Bull}\ \emph {et~al.}(2015)\citenamefont {Bull}, \citenamefont {Ferreira}, \citenamefont {Patel},\ and\ \citenamefont {Santos}}]{bull2015RadioFisher}%
  \BibitemOpen
  \bibfield  {author} {\bibinfo {author} {\bibfnamefont {P.}~\bibnamefont {Bull}}, \bibinfo {author} {\bibfnamefont {P.~G.}\ \bibnamefont {Ferreira}}, \bibinfo {author} {\bibfnamefont {P.}~\bibnamefont {Patel}}, \ and\ \bibinfo {author} {\bibfnamefont {M.~G.}\ \bibnamefont {Santos}},\ }\href@noop {} {\bibfield  {journal} {\bibinfo  {journal} {The Astrophysical Journal}\ }\textbf {\bibinfo {volume} {803}},\ \bibinfo {pages} {21} (\bibinfo {year} {2015})}\BibitemShut {NoStop}%
\bibitem [{\citenamefont {Seo}\ and\ \citenamefont {Eisenstein}(2007)}]{seo2007baoforecast}%
  \BibitemOpen
  \bibfield  {author} {\bibinfo {author} {\bibfnamefont {H.-J.}\ \bibnamefont {Seo}}\ and\ \bibinfo {author} {\bibfnamefont {D.~J.}\ \bibnamefont {Eisenstein}},\ }\href@noop {} {\bibfield  {journal} {\bibinfo  {journal} {The Astrophysical Journal}\ }\textbf {\bibinfo {volume} {665}},\ \bibinfo {pages} {14} (\bibinfo {year} {2007})}\BibitemShut {NoStop}%
\bibitem [{\citenamefont {Wu}\ and\ \citenamefont {Zhang}(2022)}]{wu2022prospects}%
  \BibitemOpen
  \bibfield  {author} {\bibinfo {author} {\bibfnamefont {P.-J.}\ \bibnamefont {Wu}}\ and\ \bibinfo {author} {\bibfnamefont {X.}~\bibnamefont {Zhang}},\ }\href@noop {} {\bibfield  {journal} {\bibinfo  {journal} {Journal of Cosmology and Astroparticle Physics}\ }\textbf {\bibinfo {volume} {2022}},\ \bibinfo {pages} {060} (\bibinfo {year} {2022})}\BibitemShut {NoStop}%
\bibitem [{\citenamefont {Battye}\ \emph {et~al.}(2013)\citenamefont {Battye}, \citenamefont {Browne}, \citenamefont {Dickinson}, \citenamefont {Heron}, \citenamefont {Maffei},\ and\ \citenamefont {Pourtsidou}}]{battye2013bingo1}%
  \BibitemOpen
  \bibfield  {author} {\bibinfo {author} {\bibfnamefont {R.}~\bibnamefont {Battye}}, \bibinfo {author} {\bibfnamefont {I.}~\bibnamefont {Browne}}, \bibinfo {author} {\bibfnamefont {C.}~\bibnamefont {Dickinson}}, \bibinfo {author} {\bibfnamefont {G.}~\bibnamefont {Heron}}, \bibinfo {author} {\bibfnamefont {B.}~\bibnamefont {Maffei}}, \ and\ \bibinfo {author} {\bibfnamefont {A.}~\bibnamefont {Pourtsidou}},\ }\href@noop {} {\bibfield  {journal} {\bibinfo  {journal} {Monthly Notices of the Royal Astronomical Society}\ }\textbf {\bibinfo {volume} {434}},\ \bibinfo {pages} {1239} (\bibinfo {year} {2013})}\BibitemShut {NoStop}%
\bibitem [{\citenamefont {Wuensche}\ \emph {et~al.}(2020)\citenamefont {Wuensche}, \citenamefont {Reitano}, \citenamefont {Peel}, \citenamefont {Browne}, \citenamefont {Maffei}, \citenamefont {Abdalla}, \citenamefont {Radcliffe}, \citenamefont {Abdalla}, \citenamefont {Barosi}, \citenamefont {Liccardo} \emph {et~al.}}]{wuensche2020bingo2}%
  \BibitemOpen
  \bibfield  {author} {\bibinfo {author} {\bibfnamefont {C.}~\bibnamefont {Wuensche}}, \bibinfo {author} {\bibfnamefont {L.}~\bibnamefont {Reitano}}, \bibinfo {author} {\bibfnamefont {M.}~\bibnamefont {Peel}}, \bibinfo {author} {\bibfnamefont {I.}~\bibnamefont {Browne}}, \bibinfo {author} {\bibfnamefont {B.}~\bibnamefont {Maffei}}, \bibinfo {author} {\bibfnamefont {E.}~\bibnamefont {Abdalla}}, \bibinfo {author} {\bibfnamefont {C.}~\bibnamefont {Radcliffe}}, \bibinfo {author} {\bibfnamefont {F.}~\bibnamefont {Abdalla}}, \bibinfo {author} {\bibfnamefont {L.}~\bibnamefont {Barosi}}, \bibinfo {author} {\bibfnamefont {V.}~\bibnamefont {Liccardo}},  \emph {et~al.},\ }\href@noop {} {\bibfield  {journal} {\bibinfo  {journal} {Experimental Astronomy}\ }\textbf {\bibinfo {volume} {50}},\ \bibinfo {pages} {125} (\bibinfo {year} {2020})}\BibitemShut {NoStop}%
\bibitem [{\citenamefont {Abdalla}\ \emph {et~al.}(2022{\natexlab{b}})\citenamefont {Abdalla}, \citenamefont {Ferreira}, \citenamefont {Landim}, \citenamefont {Costa}, \citenamefont {Fornazier}, \citenamefont {Abdalla}, \citenamefont {Barosi}, \citenamefont {Brito}, \citenamefont {Queiroz}, \citenamefont {Villela} \emph {et~al.}}]{abdalla2022bingo3}%
  \BibitemOpen
  \bibfield  {author} {\bibinfo {author} {\bibfnamefont {E.}~\bibnamefont {Abdalla}}, \bibinfo {author} {\bibfnamefont {E.~G.}\ \bibnamefont {Ferreira}}, \bibinfo {author} {\bibfnamefont {R.~G.}\ \bibnamefont {Landim}}, \bibinfo {author} {\bibfnamefont {A.~A.}\ \bibnamefont {Costa}}, \bibinfo {author} {\bibfnamefont {K.~S.}\ \bibnamefont {Fornazier}}, \bibinfo {author} {\bibfnamefont {F.~B.}\ \bibnamefont {Abdalla}}, \bibinfo {author} {\bibfnamefont {L.}~\bibnamefont {Barosi}}, \bibinfo {author} {\bibfnamefont {F.~A.}\ \bibnamefont {Brito}}, \bibinfo {author} {\bibfnamefont {A.~R.}\ \bibnamefont {Queiroz}}, \bibinfo {author} {\bibfnamefont {T.}~\bibnamefont {Villela}},  \emph {et~al.},\ }\href@noop {} {\bibfield  {journal} {\bibinfo  {journal} {Astronomy \& Astrophysics}\ }\textbf {\bibinfo {volume} {664}},\ \bibinfo {pages} {A14} (\bibinfo {year} {2022}{\natexlab{b}})}\BibitemShut {NoStop}%
\bibitem [{\citenamefont {Nan}\ \emph {et~al.}(2011)\citenamefont {Nan}, \citenamefont {Li}, \citenamefont {Jin}, \citenamefont {Wang}, \citenamefont {Zhu}, \citenamefont {Zhu}, \citenamefont {Zhang}, \citenamefont {Yue},\ and\ \citenamefont {Qian}}]{nan2011fast1}%
  \BibitemOpen
  \bibfield  {author} {\bibinfo {author} {\bibfnamefont {R.}~\bibnamefont {Nan}}, \bibinfo {author} {\bibfnamefont {D.}~\bibnamefont {Li}}, \bibinfo {author} {\bibfnamefont {C.}~\bibnamefont {Jin}}, \bibinfo {author} {\bibfnamefont {Q.}~\bibnamefont {Wang}}, \bibinfo {author} {\bibfnamefont {L.}~\bibnamefont {Zhu}}, \bibinfo {author} {\bibfnamefont {W.}~\bibnamefont {Zhu}}, \bibinfo {author} {\bibfnamefont {H.}~\bibnamefont {Zhang}}, \bibinfo {author} {\bibfnamefont {Y.}~\bibnamefont {Yue}}, \ and\ \bibinfo {author} {\bibfnamefont {L.}~\bibnamefont {Qian}},\ }\href@noop {} {\bibfield  {journal} {\bibinfo  {journal} {International Journal of Modern Physics D}\ }\textbf {\bibinfo {volume} {20}},\ \bibinfo {pages} {989} (\bibinfo {year} {2011})}\BibitemShut {NoStop}%
\bibitem [{\citenamefont {Smoot}\ and\ \citenamefont {Debono}(2017)}]{smoot2017fast2}%
  \BibitemOpen
  \bibfield  {author} {\bibinfo {author} {\bibfnamefont {G.~F.}\ \bibnamefont {Smoot}}\ and\ \bibinfo {author} {\bibfnamefont {I.}~\bibnamefont {Debono}},\ }\href@noop {} {\bibfield  {journal} {\bibinfo  {journal} {Astronomy \& Astrophysics}\ }\textbf {\bibinfo {volume} {597}},\ \bibinfo {pages} {A136} (\bibinfo {year} {2017})}\BibitemShut {NoStop}%
\bibitem [{\citenamefont {Jonas}(2009)}]{jonas2009meerkat1}%
  \BibitemOpen
  \bibfield  {author} {\bibinfo {author} {\bibfnamefont {J.~L.}\ \bibnamefont {Jonas}},\ }\href@noop {} {\bibfield  {journal} {\bibinfo  {journal} {Proceedings of the IEEE}\ }\textbf {\bibinfo {volume} {97}},\ \bibinfo {pages} {1522} (\bibinfo {year} {2009})}\BibitemShut {NoStop}%
\bibitem [{\citenamefont {Santos}\ \emph {et~al.}(2015)\citenamefont {Santos}, \citenamefont {Bull}, \citenamefont {Alonso}, \citenamefont {Camera}, \citenamefont {Ferreira}, \citenamefont {Bernardi}, \citenamefont {Maartens}, \citenamefont {Viel}, \citenamefont {Villaescusa-Navarro}, \citenamefont {Abdalla} \emph {et~al.}}]{santos2015meerkat2}%
  \BibitemOpen
  \bibfield  {author} {\bibinfo {author} {\bibfnamefont {M.~G.}\ \bibnamefont {Santos}}, \bibinfo {author} {\bibfnamefont {P.}~\bibnamefont {Bull}}, \bibinfo {author} {\bibfnamefont {D.}~\bibnamefont {Alonso}}, \bibinfo {author} {\bibfnamefont {S.}~\bibnamefont {Camera}}, \bibinfo {author} {\bibfnamefont {P.~G.}\ \bibnamefont {Ferreira}}, \bibinfo {author} {\bibfnamefont {G.}~\bibnamefont {Bernardi}}, \bibinfo {author} {\bibfnamefont {R.}~\bibnamefont {Maartens}}, \bibinfo {author} {\bibfnamefont {M.}~\bibnamefont {Viel}}, \bibinfo {author} {\bibfnamefont {F.}~\bibnamefont {Villaescusa-Navarro}}, \bibinfo {author} {\bibfnamefont {F.~B.}\ \bibnamefont {Abdalla}},  \emph {et~al.},\ }\href@noop {} {\bibfield  {journal} {\bibinfo  {journal} {arXiv preprint arXiv:1501.03989}\ } (\bibinfo {year} {2015})}\BibitemShut {NoStop}%
\bibitem [{\citenamefont {Bacon}\ \emph {et~al.}(2020)\citenamefont {Bacon}, \citenamefont {Battye}, \citenamefont {Bull}, \citenamefont {Camera}, \citenamefont {Ferreira}, \citenamefont {Harrison}, \citenamefont {Parkinson}, \citenamefont {Pourtsidou}, \citenamefont {Santos}, \citenamefont {Wolz} \emph {et~al.}}]{bacon2020meerkat3}%
  \BibitemOpen
  \bibfield  {author} {\bibinfo {author} {\bibfnamefont {D.~J.}\ \bibnamefont {Bacon}}, \bibinfo {author} {\bibfnamefont {R.~A.}\ \bibnamefont {Battye}}, \bibinfo {author} {\bibfnamefont {P.}~\bibnamefont {Bull}}, \bibinfo {author} {\bibfnamefont {S.}~\bibnamefont {Camera}}, \bibinfo {author} {\bibfnamefont {P.~G.}\ \bibnamefont {Ferreira}}, \bibinfo {author} {\bibfnamefont {I.}~\bibnamefont {Harrison}}, \bibinfo {author} {\bibfnamefont {D.}~\bibnamefont {Parkinson}}, \bibinfo {author} {\bibfnamefont {A.}~\bibnamefont {Pourtsidou}}, \bibinfo {author} {\bibfnamefont {M.~G.}\ \bibnamefont {Santos}}, \bibinfo {author} {\bibfnamefont {L.}~\bibnamefont {Wolz}},  \emph {et~al.},\ }\href@noop {} {\bibfield  {journal} {\bibinfo  {journal} {Publications of the Astronomical Society of Australia}\ }\textbf {\bibinfo {volume} {37}},\ \bibinfo {pages} {e007} (\bibinfo {year} {2020})}\BibitemShut {NoStop}%
\bibitem [{\citenamefont {Dewdney}\ \emph {et~al.}(2013)\citenamefont {Dewdney}, \citenamefont {Turner}, \citenamefont {Millenaar}, \citenamefont {McCool}, \citenamefont {Lazio},\ and\ \citenamefont {Cornwell}}]{dewdney2013ska}%
  \BibitemOpen
  \bibfield  {author} {\bibinfo {author} {\bibfnamefont {P.}~\bibnamefont {Dewdney}}, \bibinfo {author} {\bibfnamefont {W.}~\bibnamefont {Turner}}, \bibinfo {author} {\bibfnamefont {R.}~\bibnamefont {Millenaar}}, \bibinfo {author} {\bibfnamefont {R.}~\bibnamefont {McCool}}, \bibinfo {author} {\bibfnamefont {J.}~\bibnamefont {Lazio}}, \ and\ \bibinfo {author} {\bibfnamefont {T.}~\bibnamefont {Cornwell}},\ }\href@noop {} {\bibfield  {journal} {\bibinfo  {journal} {Document number SKA-TEL-SKO-DD-001 Revision}\ }\textbf {\bibinfo {volume} {1}} (\bibinfo {year} {2013})}\BibitemShut {NoStop}%
\bibitem [{\citenamefont {Newburgh}\ \emph {et~al.}(2014)\citenamefont {Newburgh}, \citenamefont {Addison}, \citenamefont {Amiri}, \citenamefont {Bandura}, \citenamefont {Bond}, \citenamefont {Connor}, \citenamefont {Cliche}, \citenamefont {Davis}, \citenamefont {Deng}, \citenamefont {Denman} \emph {et~al.}}]{newburgh2014chime1}%
  \BibitemOpen
  \bibfield  {author} {\bibinfo {author} {\bibfnamefont {L.~B.}\ \bibnamefont {Newburgh}}, \bibinfo {author} {\bibfnamefont {G.~E.}\ \bibnamefont {Addison}}, \bibinfo {author} {\bibfnamefont {M.}~\bibnamefont {Amiri}}, \bibinfo {author} {\bibfnamefont {K.}~\bibnamefont {Bandura}}, \bibinfo {author} {\bibfnamefont {J.~R.}\ \bibnamefont {Bond}}, \bibinfo {author} {\bibfnamefont {L.}~\bibnamefont {Connor}}, \bibinfo {author} {\bibfnamefont {J.-F.}\ \bibnamefont {Cliche}}, \bibinfo {author} {\bibfnamefont {G.}~\bibnamefont {Davis}}, \bibinfo {author} {\bibfnamefont {M.}~\bibnamefont {Deng}}, \bibinfo {author} {\bibfnamefont {N.}~\bibnamefont {Denman}},  \emph {et~al.},\ }in\ \href@noop {} {\emph {\bibinfo {booktitle} {Ground-based and Airborne Telescopes V}}},\ Vol.\ \bibinfo {volume} {9145}\ (\bibinfo {organization} {SPIE},\ \bibinfo {year} {2014})\ pp.\ \bibinfo {pages} {1709--1726}\BibitemShut {NoStop}%
\bibitem [{\citenamefont {Bandura}\ \emph {et~al.}(2014)\citenamefont {Bandura}, \citenamefont {Addison}, \citenamefont {Amiri}, \citenamefont {Bond}, \citenamefont {Campbell-Wilson}, \citenamefont {Connor}, \citenamefont {Cliche}, \citenamefont {Davis}, \citenamefont {Deng}, \citenamefont {Denman} \emph {et~al.}}]{bandura2014chime2}%
  \BibitemOpen
  \bibfield  {author} {\bibinfo {author} {\bibfnamefont {K.}~\bibnamefont {Bandura}}, \bibinfo {author} {\bibfnamefont {G.~E.}\ \bibnamefont {Addison}}, \bibinfo {author} {\bibfnamefont {M.}~\bibnamefont {Amiri}}, \bibinfo {author} {\bibfnamefont {J.~R.}\ \bibnamefont {Bond}}, \bibinfo {author} {\bibfnamefont {D.}~\bibnamefont {Campbell-Wilson}}, \bibinfo {author} {\bibfnamefont {L.}~\bibnamefont {Connor}}, \bibinfo {author} {\bibfnamefont {J.-F.}\ \bibnamefont {Cliche}}, \bibinfo {author} {\bibfnamefont {G.}~\bibnamefont {Davis}}, \bibinfo {author} {\bibfnamefont {M.}~\bibnamefont {Deng}}, \bibinfo {author} {\bibfnamefont {N.}~\bibnamefont {Denman}},  \emph {et~al.},\ }in\ \href@noop {} {\emph {\bibinfo {booktitle} {Ground-based and Airborne Telescopes V}}},\ Vol.\ \bibinfo {volume} {9145}\ (\bibinfo {organization} {SPIE},\ \bibinfo {year} {2014})\ pp.\ \bibinfo {pages} {738--757}\BibitemShut {NoStop}%
\bibitem [{\citenamefont {Chen}(2012)}]{chen2012tianlai1}%
  \BibitemOpen
  \bibfield  {author} {\bibinfo {author} {\bibfnamefont {X.}~\bibnamefont {Chen}},\ }in\ \href@noop {} {\emph {\bibinfo {booktitle} {International Journal of Modern Physics: Conference Series}}},\ Vol.~\bibinfo {volume} {12}\ (\bibinfo {organization} {World Scientific},\ \bibinfo {year} {2012})\ pp.\ \bibinfo {pages} {256--263}\BibitemShut {NoStop}%
\bibitem [{\citenamefont {Li}\ \emph {et~al.}(2020)\citenamefont {Li}, \citenamefont {Zuo}, \citenamefont {Wu}, \citenamefont {Wang}, \citenamefont {Zhang}, \citenamefont {Sun}, \citenamefont {Xu}, \citenamefont {Yu}, \citenamefont {Ansari}, \citenamefont {Li} \emph {et~al.}}]{li2020tianlai2}%
  \BibitemOpen
  \bibfield  {author} {\bibinfo {author} {\bibfnamefont {J.}~\bibnamefont {Li}}, \bibinfo {author} {\bibfnamefont {S.}~\bibnamefont {Zuo}}, \bibinfo {author} {\bibfnamefont {F.}~\bibnamefont {Wu}}, \bibinfo {author} {\bibfnamefont {Y.}~\bibnamefont {Wang}}, \bibinfo {author} {\bibfnamefont {J.}~\bibnamefont {Zhang}}, \bibinfo {author} {\bibfnamefont {S.}~\bibnamefont {Sun}}, \bibinfo {author} {\bibfnamefont {Y.}~\bibnamefont {Xu}}, \bibinfo {author} {\bibfnamefont {Z.}~\bibnamefont {Yu}}, \bibinfo {author} {\bibfnamefont {R.}~\bibnamefont {Ansari}}, \bibinfo {author} {\bibfnamefont {Y.}~\bibnamefont {Li}},  \emph {et~al.},\ }\href@noop {} {\bibfield  {journal} {\bibinfo  {journal} {SCIENCE CHINA Physics, Mechanics \& Astronomy}\ }\textbf {\bibinfo {volume} {63}},\ \bibinfo {pages} {129862} (\bibinfo {year} {2020})}\BibitemShut {NoStop}%
\end{thebibliography}%

\end{document}